**Tailoring phonon-driven responses in α-MoO$_3$ through isotopic enrichment**

Thiago S. Arnaud[1,2], Ryan W. Spangler[3], Johnathan D. Georgaras[4], Jonah B. Haber[4], Daniel Hirt[5], Maximilian Obst[1,2,6,7], Gonzalo Álvarez-Pérez[8,9], Mackey Long III[1,2], Felix G. Kaps[6,7], Jakob Wetzel[6,7], Courtney Ragle[1,2], John E. Buchner[1,2], Youngji Kim[2], Aditha S. Senarath[1,2], Richarda Niemann[2], Mingze He[2], Giulia Carini[9], Unai Arregui-Leon[9,10], Akash C. Behera[9], Ramachandra Bangari[12], Nihar Sahoo[12], Niels C. Brumby[9], J. Michael Klopf[11], Martin Wolf[9], Lukas M. Eng[6,7], Susanne C. Kehr[6,7], Thomas G. Folland[12], Alexander Paarmann[9], Patrick E. Hopkins[5,13,14], Felipe Jornada[4]*, Jon-Paul Maria[3]*, Joshua D. Caldwell[1,2]*

[1] Interdisciplinary Material Science, Vanderbilt University, Nashville 37235, TN, USA

[2] Department of Mechanical Engineering, Vanderbilt University, Nashville 37235, TN, USA

[3] Department of Materials Science and Engineering, The Pennsylvania State University, University Park 16802, PA, USA

[4] Department of Materials Science and Engineering, Stanford University, Stanford, CA 94305, USA

[5] Department of Mechanical and Aerospace Engineering, University of Virginia, Charlottesville, Virginia 22904, USA

[6] Institute of Applied Physics, TUD Dresden University of Technology, Dresden 01187, Germany

[7] Würzburg-Dresden Cluster of Excellence - EXC 2147 (ct.qmat), Dresden 01062, Germany

[8] Istituto Italiano di Tecnologia, Center for Biomolecular Nanotechnologies, Lecce, Italy

[9] Department of Physcial Chemistry, Fritz Haber Institute of the Max Planck Society, Faradayweg 4-6, 14195, Berlin, Germany

[10] Department of Physics, Politecnico di Milano, Piazza Leonardo da Vinci 32, 20133, Milan, Italy

[11] Institute of Radiation Physics, Helmholtz-Zentrum Dresden-Rossendorf, Dresden 01328, Germany

[12] Department of Physics and Astronomy, University of Iowa, Iowa City 52242, IA, USA

[13] Department of Materials Science and Engineering, University of Virginia, Charlottesville, Virginia 22904, USA

[14] Department of Physics, University of Virginia, Charlottesville, Virginia 22904, USA

* Correspondence to: josh.caldwell@vanderbilt.edu, jpm133@psu.edu, and jornada@stanford.edu

**Abstract**

The implementation of polaritonic materials into nanoscale devices requires selective tuning of parameters to realize desired spectral or thermal responses. One robust material is α-MoO$_3$, which as an orthorhombic crystal boasts three distinct phonon dispersions, providing three polaritonic dispersions of hyperbolic phonon polaritons (HPhPs) across the mid-infrared (MIR). Here, the tunability of both optical and thermal responses in isotopically enriched α-MoO$_3$ ($^{98}$MoO$_3$, Mo$^{18}$O$_3$ and $^{98}$Mo$^{18}$O$_3$) are explored. A uniform ~5% spectral redshift from $^{18}$O enrichment is observed in both Raman- and IR-active TO phonons. Both the in- and out-of-plane thermal conductivities for the isotopic variations are reported. *Ab initio* calculations both replicate experimental findings and analyze the select-mode three-phonon scattering contributions. The HPhPs from each isotopic variation are probed with s-SNOM and their *Q*-factors are reported. A *Q*-factor maxima increase of ~50% along the [100] in the RB$_2$ and ~100% along the [001] in the RB$_3$ are reported for HPhPs supported in $^{98}$Mo$^{18}$O$_3$. Observations in both real and Fourier space of higher-order HPhP modes propagating in single slabs of isotopically enriched α-MoO$_3$ without the use of a subdiffractional surface scatterer are presented here. This work illustrates the tunability of α-MoO$_3$ for thermal and nanophotonic applications.

**Introduction**

Polaritonic materials offer the subwavelength, directional, nanoscale confinement of light via polaritons – quasiparticles of light and matter that transport both energy and phase. Phonon polaritons are supported when a photon is coupled to the dipolar lattice oscillations (i.e. optical phonons) at a resonant frequency[1]. This resonant frequency is within the spectral regime known as the Reststrahlen band (RB), splitting between the transverse (TO) and longitudinal optical (LO) phonons, where the real parts of the dielectric permittivity tensor are negative.

Hyperbolic phonon polaritons (HPhPs) are a class of phonon polaritons with ray-like propagation throughout the bulk of the crystal. As the name implies, HPhPs are supported within uniaxial (biaxial) hyperbolic crystals with spectral regions where at least one principal axes features a real part of the permittivity tensor that is negative, while at least one is positive[2]. HPhPs have been extensively studied in many uniaxial crystals, with the initial observations reported within hexagonal boron nitride[3,4] (hBN), through launching mechanisms[5,6], manipulation of substrate refractive index[7], hybrid modes within heterostructures[8,9], modulation via highly doped semiconductor substrates[10], and hyperlensing[11–14]. Molybdenum trioxide (α-MoO$_3$) is an orthorhombic van der Waals crystal with natural biaxial hyperbolicity that elicits anisotropic in-plane HPhP propagation within the three Reststrahlen bands (RB$_{1-3}$) in the MIR and terahertz regime that each correspond to one of the three principal crystal directions[15–17]. Therefore, MoO$_3$ intrinsically is capable of guiding in-plane HPhPs between asymptotic limits along one crystallographic direction, while simultaneously restricting propagation in the others, leading to promise for applications in mid IR polarizers[18], polaritonic-driven chemistry[19], in-plane HPhP focusing[20–23], nanoresonators based on negative reflection[24], negative refraction[25], geometric confinement[26], substrate mediation[27,28], twist-optics[9,29–36], chiral emission[37], hypercrystals[38], and gate tuning HPhPs coupled with graphene plasmons[39–41].

Aside from tuning HPhPs through the external environment[7], isotopic enrichments offer an avenue to intrinsically control polariton lifetimes and dispersion. *Caldwell et al.* discussed the significance of isotope enrichment in a diatomic crystal, where the acoustic phonon lifetimes are primarily dictated by the heavier mass and the optic phonon lifetimes are mostly dependent on the lighter mass.[1] Eliminating the isotopic disorder present in naturally abundant crystals inherently lowers the HPhP loss through reduced scattering rates. *Giles et al.* demonstrated a 3-4 fold improvement in the optic phonon lifetime in isotopically enriched hBN, which translated into corresponding improved HPhP lifetimes and propagation lengths.[42] While the optic phonon lifetime improvements were shown through the low-loss HPhPs supported within isotopically enriched hBN, the small differences in atomic mass between boron and nitrogen makes it difficult to discern the atomic dependence between optic and acoustic modes. A recent study performed by *Carini et al.* observed a ~40 cm$^{-1}$ redshift in phonon frequencies in monoclinic β-Ga$_2$O$_3$ (bGO) when isotopically substituting $^{16}$O with $^{18}$O.[43] Similar to bGO, α-MoO$_3$ also has a large atomic mass difference, which serves as an excellent platform to unveil specific spectral responses and modifications in the thermal response via selective isotopic enrichment.

Naturally abundant MoO$_3$ (nat-MoO$_3$) is comprised of seven stable Mo isotopes with significant concentrations: $^{92}$Mo (15.86%), $^{94}$Mo (9.12%), $^{95}$Mo (15.70%), $^{96}$Mo (16.50%), $^{97}$Mo (9.45%), $^{98}$Mo (23.75%), and $^{100}$Mo (9.62%)[44]. *Zhao et al.* studied HPhPs within RB$_3$ supported along the [100] in isotopically pure $^{92}$MoO$_3$ and $^{100}$MoO$_3$ and reported improved propagation lengths and lifetimes for both isotopes in comparison to nat-MoO$_3$.[45] *Schultz et al.* later performed a similar investigation of the same Mo isotopes within RB$_2$, realizing modest improvements in propagation lengths and lifetimes.[46] However, a comprehensive study has yet to experimentally investigate the individual atomic mass dependences in α-MoO$_3$ and differentiate their respective influences upon the acoustic and optic phonon-driven responses.

Here we reveal the anisotropic losses present in the MIR optical phonons of α-MoO$_3$ via isotopic enrichment of both oxygen and molybdenum with respect to natural abundant α-MoO$_3$. We prepare samples of varying isotopic content: $^{98}$MoO$_3$, Mo$^{18}$O$_3$ and $^{98}$Mo$^{18}$O$_3$ to investigate implications of a ~12.5% mass increase in the lighter mass and further study the role of the reduced Mo isotopic disorder plays in the optical and thermal properties. In the far-field regime, we observe significant phonon frequency red shifts due to $^{18}$O enrichment from Raman and Fourier transform infrared (FTIR) spectroscopy. Regarding the thermal properties, we utilize time-domain thermoreflectance (TDTR) to extract the thermal conductivity from each isotopic sample along the in- and through-plane directions. We find good agreement between first-principles density functional perturbation theory (DFPT) calculations and the thermal and far-field optical experimental results. We provide additional DFT analysis on the acoustic and optical phonon scattering processes governed by their respective scattering phase-space and the corresponding mode-dependent coupling strength (set by their eigenvectors and degree of Mo-O atomic character). Shifting to the near-field regime, we probe HPhPs in all three MIR RBs to fully characterize the polaritonic nature of isotopically enriched α-MoO$_3$. Notable findings include relative increases in HPhP $Q$-factor with respect to the phonon damping in the RB$_2$ and RB$_3$. Additionally, we found that, within the RB$_3$, the in-plane direction that supports higher $Q$ HPhPs flip when isotopically enriched. Furthermore, we report the direct, real-space imaging of higher order modes in single slabs of isotopic α-MoO$_3$. This work investigates the interplay between Mo and O isotopic enrichment in α-MoO$_3$, which enables both the spectral tuning of phonon frequencies and the in-plane anisotropic enhancement of HPhP $Q$-factors to enhance its applications in MIR polarizers, chiral emission, and nanoscale confinement and manipulation of both light and heat.

**Results**

Reactive vapor transport was used to grow α-MoO$_3$ crystals of varying isotopic contents (see methods): naturally abundant α-MoO$_3$, $^{98}$MoO$_3$, Mo$^{18}$O$_3$, and $^{98}$Mo$^{18}$O$_3$. The $^{98}$Mo isotope was selected out of the seven stable options due to it being the largest quantity present in natural abundant α-MoO$_3$ with the intent of establishing an enriched sample closest to nat-MoO$_3$ without the isotopic disorder scattering loss or additional spectral shifts contributed from other Mo isotopes. The near monoisotopic crystals were then characterized using Raman spectroscopy (see methods), where a systematic ~5% red shift from nat-MoO$_3$ was observed in both Mo$^{18}$O$_3$ and $^{98}$Mo$^{18}$O$_3$ (**Figure 1a**). It is important to note that the percent red shift from oxygen enrichment does not hold for acoustic phonons below ~190 cm$^{-1}$ as seen in the DFPT calculated phonon dispersion (**Figure S1.1**; see methods). These shifts in the higher energy phonons are in agreement with reported Raman shifts from $^{18}$O enriched bGO.[47] Beyond the simple difference in relative mass increase (~12.5% for O vs. ~2% for Mo), we identify that these MIR optic phonons are predominantly oxygen in character (see **Figure S1.2**), making them exceptionally sensitive to detuning via oxygen substitution. Comparison between the measured and *ab initio* linewidths of the TO Raman peaks reveal excellent agreement between expected and observed linewidth reductions from $^{18}$O enrichment (**Figure 1b-d**). The full dependence of these mechanisms on optical mode character and atomic participation is further discussed in **Section S2**. We note that the apparent discrepancies between the experimental and theoretical values in **Figure 1d** — particularly for the narrow B$_{2g}$ mode — is a result of the experimental spectral resolution limit of 0.8 cm$^{-1}$ (see methods). Since the predicted intrinsic anharmonic linewidth of the $B_{2g}$ mode ($\approx 0.66\,\text{cm}^{-1}$) falls below this instrumental threshold, the experimental measurements are instrument-limited and cannot fully resolve the fine linewidth narrowing predicted by theory. However, the theoretical trends remain consistent with the physical picture of moderate lifetime improvements from $^{18}O$ enrichment.

To correlate our far-field Raman measurements and make appropriate near-field approximations, we characterize the IR-active phonons through FTIR spectroscopy (see methods and **Section S3** for more details). Reflection spectra were collected at near-normal incidence polarized along the [100] and [001] directions, showing the characteristic high reflectivity Reststrahlen bands of the $RB_1$ and $RB_2$ plotted in **Figure S3.2a**, respectively. Due to the angular spread present in the reflective optics in the measurements, we observe additional dips in the reflectivity that correspond to the out-of-plane $RB_3$ TO phonon and additional modes further discussed from **Figure S3.2b**. Additional measurements were taken at an oblique angle of incidence to increase sensitivity to the out-of-plane $RB_3$. (**Figure S3.2c-d**) Similar to the Raman spectra, redshifts of ~5% in the IR-active TO phonons are observed due to isotopic enrichment of $^{18}O$ in α-$MoO_3$. As previously mentioned, near equivalent shifts in the RBs were observed in $^{18}O$ enriched bGO.[43] Such agreement in $^{18}O$ studies in different crystals clearly supports the idea that the optic phonons are strongly governed by the lighter mass in a diatomic crystal.

Rather than identifying the scattering channels contributing to each TO phonon linewidth, as done with the Raman modes in Section S2, we focus on the *change* in three-phonon scattering contributions between $MoO_3$ and $Mo^{18}O_3$. Thus, we investigate the microscopic origin of the linewidth narrowing under $^{18}O$ enrichment via the linewidth difference histograms for the $RB_{1-3}$ TO phonons (parent-phonon frequencies) shown in **Figure 1e-g**. The difference maps report the change in the mode-resolved three-phonon contribution to the intrinsic linewidth when isotopically substituting from $^{16}O$ to $^{18}O$. The axes represent the daughter-phonon frequency pairs ($\omega_1$, $\omega_2$) that participate in energy- and momentum-conserving decay and coalescence processes. Consistent with the measured narrowed linewidths, all three modes exhibit a net reduction in intrinsic linewidth under $^{18}O$ enrichment ($RB_{1-3}$ TO: $\Delta\Gamma_{1-3}$ = -2.66, -1.16, and -0.114 cm$^{-1}$, corresponding to -11.8%, -15.0%, and -11.5%, respectively), accompanied by substantial redshifts of the parent frequencies ($\Delta\omega_0 \approx$ -25 to -42 cm$^{-1}$).

The phonon-phonon scattering rate inherits a $1/\sqrt{m}$ mass factors from eigenvector normalization, so heavier isotope substitution might naively be expected to uniformly reduce all anharmonic decay rates.[48] However, the difference histograms contain both positive and negative regions, indicating that $^{18}O$ substitution redistributes the three-phonon scattering rather than uniformly suppressing it or simply redshifting the available energy-conserving phase-space. The net linewidth reduction from $^{16}O$ to $^{18}O$ arises from an incomplete cancellation between the suppressed channels (negative regions) and enhanced channels (positive regions), with the former outweighing the latter. The mode-dependent redistribution is seen in the difference between linewidth changes of the $RB_{1-3}$ TO phonons: the linewidth decrease of the $RB_1$ TO phonon is dominated by reduced contributions involving intermediate daughter frequencies ($\omega \leq 500\,\text{cm}^{-1}$); the $RB_2$ TO phonon shows pronounced suppression for mixed low- and high-frequency daughter pairs; and the $RB_3$ TO phonon exhibits a characteristic transfer of scattering weight away from the upper edge of the phonon spectrum ($\omega \geq 1000\,\text{cm}^{-1}$) toward slightly lower daughter frequencies. These mode-dependent trends arise from non-uniform softening of the phonon spectrum, where the frequencies of different branches shift by different amounts under $^{18}O$ enrichment, reshaping the energy-conserving phase-space and redistributing the dominant three-phonon pathways.

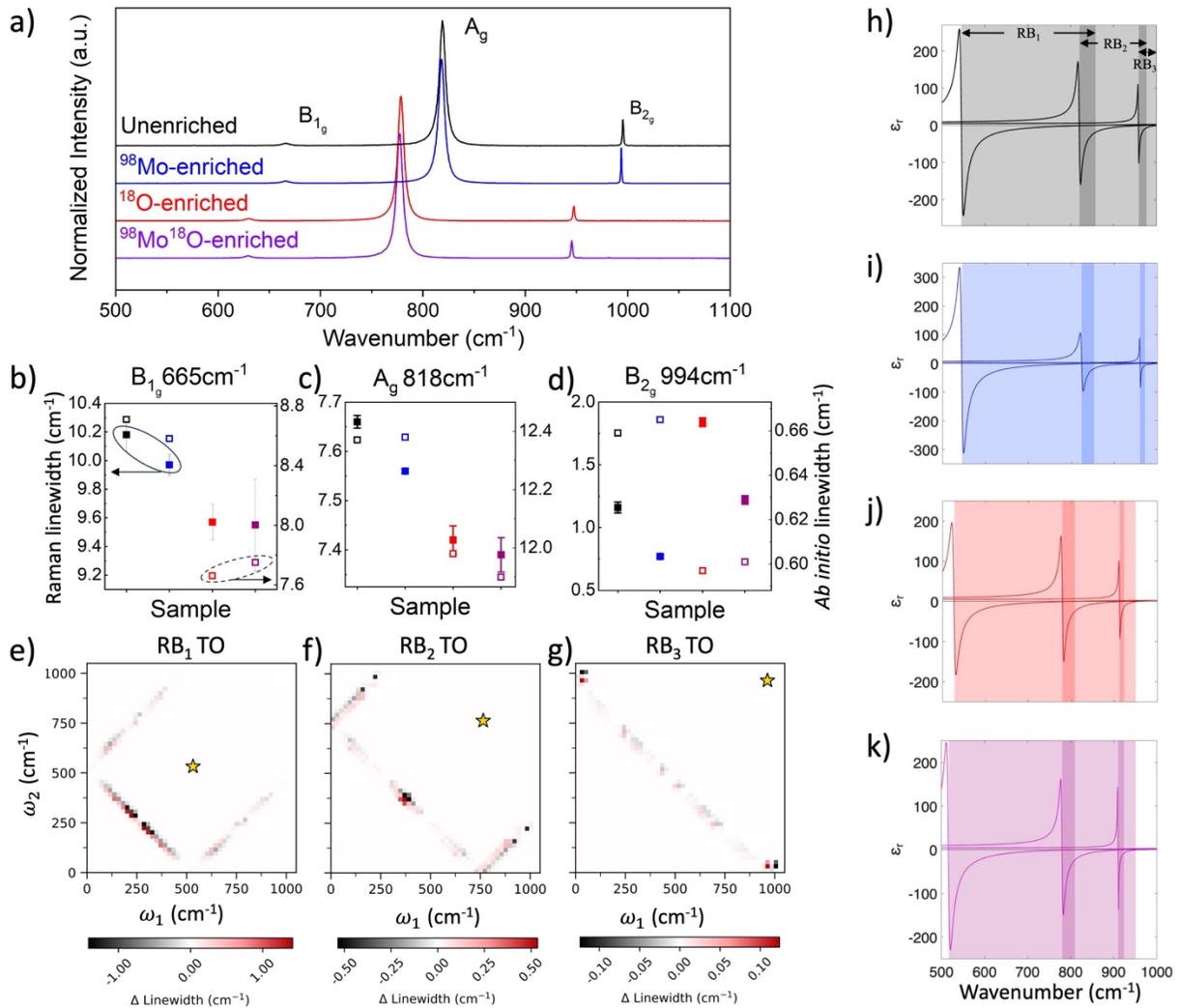

**Figure 1.** Far-field optical characterization on varying isotopic contents of α-MoO$_3$. **a)** Raman spectroscopy for each labeled isotopically enriched sample. The Raman modes are labeled beside their respective peaks. The same isotope-color scheme in part a) is repeated throughout the rest of Figure 1. **b-d)** Extracted MIR Raman linewidths from the modes labeled in a) and their frequencies with respect to naturally abundant (unenriched) α-MoO$_3$. The horizontal axis is indexed by sample for clarity since the peaks would be tightly grouped in frequency for crystals of $^{16}$O and $^{18}$O content. The error bars are ±1 standard deviation. **e-g)** Mode-resolved *difference histograms* showing the enrichment-induced redistribution of the three-phonon contributions to the intrinsic linewidth upon $^{18}$O substitution for the corresponding Γ-point modes (RB$_{1-3}$ TO phonons marked by the gold star). The axes represent the frequencies of the daughter phonon pairs ($\omega_1$, $\omega_2$) involved in the three-phonon decay and coalescence processes. Each ($\omega_1$, $\omega_2$) bin aggregates the summed three-phonon linewidth contributions from all momentum/branch contributions whose daughter frequencies fall within that bin. The color scale indicates the change of the linewidth contributions in cm$^{-1}$: black (negative) regions correspond to reduced contributions under $^{18}$O enrichment , while red (positive) regions indicate increased contributions. **h-k)** Experimentally approximated real-part of the dielectric functions for each isotopic variation of α-MoO$_3$. The parameters for the dielectric function of naturally abundant α-MoO$_3$ are taken from reported values.[15]

The IR optical characteristics of α-MoO$_3$ are captured in its dielectric function and, in the absence of free carriers, arise mainly from the various IR phonon resonances. The dielectric function for TO and LO splittings is described by a Lorentz oscillator for each principal axis in the diagonal permittivity tensor[15]:

$$\varepsilon_j(\omega) = \varepsilon_j^\infty \left( \frac{(\omega_j^{LO})^2 - \omega^2 - i\gamma_j\omega}{(\omega_j^{TO})^2 - \omega^2 - i\gamma_j\omega} \right), j = x, y, z \qquad [1]$$

where $\varepsilon_j(\omega)$ is the permittivity component denoted by *j* along the [100], [001], and [010] for x, y, and z, respectively, $\varepsilon_j^\infty$ is the high-frequency dielectric constant, $\omega_j^{TO}$ and $\omega_j^{LO}$ are the TO and LO phonon frequencies, and $\gamma_j$ is the phonon damping.

Since this study investigates isotopically enriched α-MoO$_3$ not previously reported, we provide their respective optical behaviors here. As a first approximation of the dielectric function for each isotopic sample, we alter the TO and LO phonons from naturally abundant α-MoO$_3$ according to their respective Raman and IR spectral shifts observed. Knowing the TO and LO frequencies from experimental data and approximate shifts, we utilize the Lyddane-Sachs-Teller relation[49]:

$$\frac{\varepsilon_0}{\varepsilon_j^\infty} = \left( \frac{(\omega_j^{LO})^2}{(\omega_j^{TO})^2} \right), j = x, y, z \qquad [2]$$

to estimate the $\varepsilon_j^\infty$ value for each direction with the known TO and LO frequencies and the dielectric constant, $\varepsilon_0$. The phonon damping for each isotopically enriched sample is fixed with reported values of naturally abundant α-MoO$_3$ values[15] due to a recent study from *Nandanwar et al.*,[50] which revealed overestimates in phonon damping from isotopic hBN values reported by *Giles et al.*[42] for measurements at room temperature. In consideration of all these factors, we insert the redshifted TO and LO parameters into **Equation 1** and plot the approximated dielectric functions for each isotopically enriched α-MoO$_3$ in **Figure 1h-k.** While $\gamma$ is kept at the naturally abundant α-MoO$_3$ value and therefore, not plotted, we can infer from the Raman linewidth reductions and DFPT difference scattering phase-space that the IR-active TO linewidths also experience similar reductions from $^{18}$O enrichment. These conclusions are further supported by DFPT calculated TO linewidths (**Figure S4.1**). Therefore, we expect modest phonon lifetime improvements in the $^{18}$O flakes but, it is difficult to discern between phonon-phonon scattering and additional scattering losses arising from isotopic disorder.

With strong experimental evidence indicating the far-field optic phonon characteristics dependent on the lighter of the two masses, we also examine the influence of isotopic enrichment on the lower frequency, heat carrying optic and acoustic phonon properties contingent upon the heavier mass. To this extent, we perform time-domain thermoreflectance (TDTR; see **Section S5**) measurements to extract the in- and out-of-plane thermal conductivity from each of our isotopically enriched α-MoO$_3$ flakes. The out-of-plane (in-plane) thermal conductivities on each sample are listed in **Table S5.1 (S5.4)** alongside the DFPT calculated values (see **Section S1.4** for more details). Further information is available in **Section S5** regarding necessary fittings, uncertainty analysis and thermal boundary conductance. Our measured thermal conductivities reflect the predicted thermal conductivity behavior for a vdW crystal, exhibiting higher thermal conductivity in the in- vs out-of-plane directions. We observe similar thermal conductivities across our available isotopes in this study due to little variance between the Mo atomic weight present in naturally abundant α-MoO$_3$ (~95.96 u) and the 98 u enriched flake. This equates to a ~2% change in Mo mass compared to the ~12% change in O mass from $^{16}$O to $^{18}$O. This is consistent with prior works[51], given the relative reduction in phonon scattering rates for lower frequency phonons to changes in atomic mass as compared to the higher frequency, non-thermally activated optically active phonons[52].

We further investigate the thermal conductivity of α-MoO$_3$ through *ab initio* calculations. The *ab initio* thermal conductivity values plotted in **Figure S6.1** are in great agreement with the experimental values reported in **Tables S5.1** and **S5.4** We employ the same eigenvector atomic projections onto the phonon dispersion to identify the acoustic and optic phonon contributions driven by Mo or O atoms (**Figure S6.2**). We correlate both projections upon the phonon dispersion to identify anisotropic pathways of thermal transport that can be isotopically tuned to the degree of Mo or O character. As a result, we predict a significant ~30% contribution from the optic phonons and a ~40% contribution specifically from the oxygen atoms in both the in- and cross-plane thermal transport. We breakdown these theoretical thermal conductivity predictions into their independent contributions from phonon heat capacity, phonon group velocity, phonon-phonon scattering rates, and isotope scattering (**Figure S6.3**). Thus, we predict the mechanisms behind thermal transport in α-MoO$_3$ and the influence of isotopic enrichment upon those independent contributors. Recent work has explored the ultrafast thermal transport across the Au/hBN interface mediated by the excitation of HPhPs in hBN[53]. Similar future work should explore the role of the in-plane anisotropy intrinsic to α-MoO$_3$ upon directional, nanoscale thermal transport mediated by isotopic HPhPs that is beyond the scope of this work.

In an effort to investigate the isotope influence upon the phonon damping in α-MoO$_3$, we perform near-field optical measurements of HPhPs, which are highly sensitive to their respective TO phonons to which they are coupled. We employ scattering-type scanning near-field optical microscopy (s-SNOM; see methods) of HPhPs supported in mechanically exfoliated isotopic flakes across the three MIR RBs. In this work, we utilize two different illumination sources. For the RB$_1$, the s-SNOM system is coupled with a tunable free-electron laser (FEL) to access the low frequencies not typically available in tunable table-top light sources.[54,55] The remaining RB$_2$ and RB$_3$ were probed using a tunable quantum cascade laser (QCL) as the illumination source to excite the respective in-plane hyperbolic and elliptic HPhPs. To understand the frequency dependence of the HPhPs probed in this work, we consider the free-space normalized biaxial analytical dispersion relation[15,56]

$$\frac{k}{k_o} = \frac{\rho}{k_o d}\left[\tan^{-1}\left(\frac{\rho\varepsilon_a}{\varepsilon_z}\right) + \tan^{-1}\left(\frac{\rho\varepsilon_s}{\varepsilon_z}\right) + \pi l\right], \; l = 0,1,2\ldots \quad [3]$$

and

$$\rho = i\sqrt{\frac{\varepsilon_z}{\varepsilon_x \cos^2(\alpha) + \varepsilon_y \sin^2(\alpha)}} \quad [4]$$

where $k$ is the real-valued in-plane HPhP wavevector, $k_0$ is the free space wavevector, $d$ is the flake thickness, $l$ is the discrete HPhP mode order, $\varepsilon_a$ is the permittivity of air, $\varepsilon_s$ is the permittivity of our substrate, and $\varepsilon_{x,y,z}$ are the respective in- and out-of-plane permittivities of the α-MoO$_3$. The factor $\rho$ includes the angle $\alpha$ between the x ([100]) direction and the in-plane incident excitation source wavevector- where $\alpha$ is fixed to 0° when measuring along the [100] and 90° along the [001]. The unitless polariton wavevector, $\frac{k}{k_o}$, is normalized by the free space wavevector and we only consider propagating modes ($\frac{k}{k_o} > 1$) where the HPhPs subside. s-SNOM images of HPhPs excited at similar real permittivity values in $^{98}$MoO$_3$ at 920 cm$^{-1}$ ($\varepsilon_r$=-2.22) and Mo$^{18}$O$_3$ ($\varepsilon_r$=-2.66) and $^{98}$Mo$^{18}$O$_3$ ($\varepsilon_r$=-2.71) at 880 cm$^{-1}$ are shown in **Figures 2a-c** respectively. As predicted from far-field measurements, we observe a ~40 cm$^{-1}$ red shift in the RBs from $^{18}$O enrichment since the s-SNOM images are near-identical for the offset excitation frequencies. This spectral shift is more apparent in the near-field by extracting the line profiles of these HPhPs supported in similar thickness (~200 nm) flakes of similar polariton wavelengths

($\lambda_p$) at different excitation frequencies (**Figure S7.1** for specific AFM details). The resulting line profiles from **Figure 2a-c** plotted in **Figure 2d** display a consistent $\lambda_p \sim 0.5\,\mu m$ and a 40 cm$^{-1}$ red shift from the $^{98}$MoO$_3$ at 920 cm$^{-1}$ to the Mo$^{18}$O$_3$ and $^{98}$Mo$^{18}$O$_3$ at 880 cm$^{-1}$, respectively. In all cases, both tip ($\frac{\lambda_p}{2}$) and edge-launched ($\lambda_p$) HPhPs are present due to the orientation of the flake edge set to be normal to the in-plane polarization of the incident light.

Due to the possible presence of higher-order modes in the s-SNOM measurements, we perform all our polariton analyses in momentum space. Therefore, we employed an FFT analysis (see **Section S8** for more details) to separate the HPhP modes in momentum space and extract their real (Re($k$)) and imaginary (Im($k$)) wavevectors. The HPhPs propagating in $^{98}$MoO$_3$ at 880 cm$^{-1}$ in **Figure 2e** further illustrates the 40 cm$^{-1}$ red shift when compared to the s-SNOM images of the Mo$^{18}$O$_3$ and $^{98}$Mo$^{18}$O$_3$ HPhPs in **Figures 2b** and **2c** excited at the same frequency. In addition to the comparisons of the polaritonic wavelengths outlined above, from the line scans provided, additional frequency components are apparent. These additional frequencies are clearly visible in the line profile plotted in **Figure 2f**. We identify the higher-order modes present, through a fast-Fourier transform (FFT) bandpass filter on the raw line profile to separate any noise, far-field signal, and separate the higher-order modes present in momentum space (see **Section S8** for more details). The higher order modes are designated by each distinct Fourier peak with increasing momentum. The bandpass regions are then replotted in **Figure 2f** in the respective dotted and dashed lines. It is evident that up to the possible third-order ($l = 2$) mode is present within the measured line profiles of the isotopically enriched samples. The influence of the third-order mode is primarily seen in the first fringe of the first-order ($l = 0$) mode, where the single peak is split into a doublet due to the first two fringes of the third-order occurring before the mode is estimated to decay. The second-order ($l = 1$) mode propagates further from the edge resulting in additional fringes and shoulders of the first-order mode. Prior observations of higher order modes in α-MoO$_3$ were only able to be realized by scattering incident light from a 3C-SiC nanowire[6] or by suppression of the fundamental mode using substrates with negative real permittivity such as 4H-SiC or Au[57]. Recent efforts have achieved efficient excitation of higher-order modes in α-MoO$_3$ through the implementation of a secondary scattering event.[58] Thus, this represents a direct observation of propagating, higher order modes within isotopically enriched MoO3.

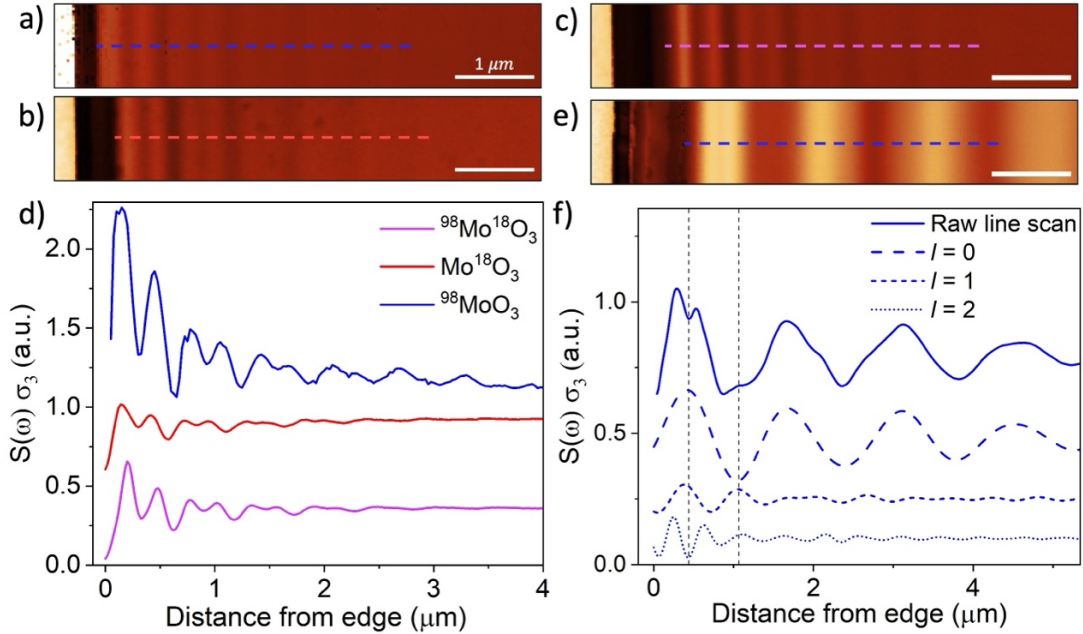

**Figure 2.** Experimental near-field characterization of varying isotopic content of α-MoO$_3$. **a-c)** Experimental SNOM measurements of similar thickness (~200 nm) $^{98}$MoO$_3$ at 920 cm$^{-1}$, Mo$^{18}$O$_3$ at 880 cm$^{-1}$, and $^{98}$Mo$^{18}$O$_3$ at 880 cm$^{-1}$, respectively. The dashed lines illustrate where the HPhP line profiles are taken from the SNOM images; correlated with AFM profiles at the start of the surface from the flake edge. **d)** Extracted line profiles from a-c) for HPhPs of similar wavelength. **e)** SNOM measurement of a 460 nm $^{98}$MoO$_3$ flake at 880 cm$^{-1}$ where higher-order modes are visibly present in the real-space image. **f)** Deconvolved FFT-filtered line profile of e) where the solid line is the raw extracted profile and subsequent profiles below are increasing in higher-order HPhP. The vertical lines are plotted as a visual aid to guide the reader in observing how the higher-order modes modulate the fundamental HPhP mode.

Using the approximate isotopic dielectric functions, we calculated the imaginary part of the reflection coefficient, Im($r_p$), with the transfer matrix method (TMM)[59] for an air/α-MoO$_3$/Si structure along both in-plane crystallographic directions. The analytical hyperbolic dispersions for $^{98}$MoO$_3$, Mo$^{18}$O$_3$, and $^{98}$Mo$^{18}$O$_3$ are overlayed on the calculated Im($r_p$) values and alongside the experimental Re($k$) values of HPhPs propagating along the [100] ([001]) axis in **Figure 3a-c (3d-f)**. We provide the dispersion mapping for different sample thicknesses in **Figure S9.1**. We see excellent agreement in all samples between the extracted HPhP momenta and their analytical dispersions. The higher order modes discussed from **Figure 2f** and present in other excitation frequencies are extracted from the FFT analysis are validated as second and third-order modes. Due to the spectral redshift from $^{18}$O enrichment, we have limited access in frequency where we can probe the third-order modes without significant damping. Thus, we demonstrate real-space imaging of reduced loss, spectrally shifted HPhPs in isotopically enriched α-MoO$_3$. Furthermore, this lower scattering loss enables the excitation and imaging of higher order HPhP modes without the aid of a sub-diffractional scattering source on the sample.

In contrast to the in-plane hyperbolic bands, the in-plane elliptical RB3 in both $^{98}$MoO3 and Mo$^{18}$O3 only supported higher-order modes along the [001] direction. In general, the narrow bandwidth of the RB3 compared to RB1-2 result in slower HPhP group velocities ($v_g = \frac{\partial \omega}{\partial k}$). It is clear from **Figure 3** that the group velocities of the RB3 are much slower along the [100] than the [001] due to the dispersion asymptotic limits for the RB3 *l=-1* branch in $^{98}$MoO3 it is roughly 961 cm$^{-1}$ along the [001] but 971 cm$^{-1}$ along the [100]. As discussed by *Sternbach et al.* in SI note 12, the spectral bandwidth of the light source imposes additional scattering losses via the excitation of similar momenta polaritons that destructively interfere with one another.[60] In a similar case, a larger range of HPhPs are excited for a slower group velocity. As a result, higher-order modes for all isotopic samples along the [100] suffer more scattering losses compared to those along the [001]. The analytical hyperbolic dispersion, see **Equation 3**, contains a thickness dependence which will result in a slower group velocity for thinner flakes. We note that the $^{98}$Mo$^{18}$O3 group velocities of the higher-order modes shown in **Figure 3f** are slower than the other two isotopic samples due to the large discrepancy in thickness. A study on the HPhP group velocity would serve as a baseline requirement for efficient excitation of higher-order modes but is beyond the scope of this work.

**Table 1:** Experimental dielectric function parameters of isotopically enriched α-MoO3

| Isotope | $\omega_x^{TO}$ (cm$^{-1}$) | $\omega_x^{LO}$ (cm$^{-1}$) | $\varepsilon_x^\infty$ | $\omega_y^{TO}$ (cm$^{-1}$) | $\omega_y^{LO}$ (cm$^{-1}$) | $\varepsilon_y^\infty$ | $\omega_z^{TO}$ (cm$^{-1}$) | $\omega_z^{LO}$ (cm$^{-1}$) | $\varepsilon_z^\infty$ |
|---|---|---|---|---|---|---|---|---|---|
| $^{98}$MoO3 | 824 | 966 | 4.3 | 544 | 849 | 7.9 | 960 | 1009 | 2.0 |
| Mo$^{18}$O3 | 779 | 923 | 5.8 | 526 | 806 | 5.1 | 912 | 954 | 4.5 |
| $^{98}$Mo$^{18}$O3 | 779 | 923 | 5.8 | 526 | 806 | 6.1 | 910 | 954 | 3.6 |

The parameters for the dielectric function describing the dispersive nature for each isotopically enriched α-MoO3, which have not been previously investigated elsewhere, are provided here. After verifying the spectral shifts in the near-field regime, we report the approximate experimental Lorentz oscillator parameters listed in **Table 1** for the isotopic α-MoO3 crystals in this work and a comparison to previously published isotopically enriched α-MoO3 values provided in **Table S10.1**. Aside from the damping parameter, the TO and LO frequencies and high-frequency dielectric constants were slightly tuned in the TMM calculations to fit with the experimental dispersion points in **Figure 3**.

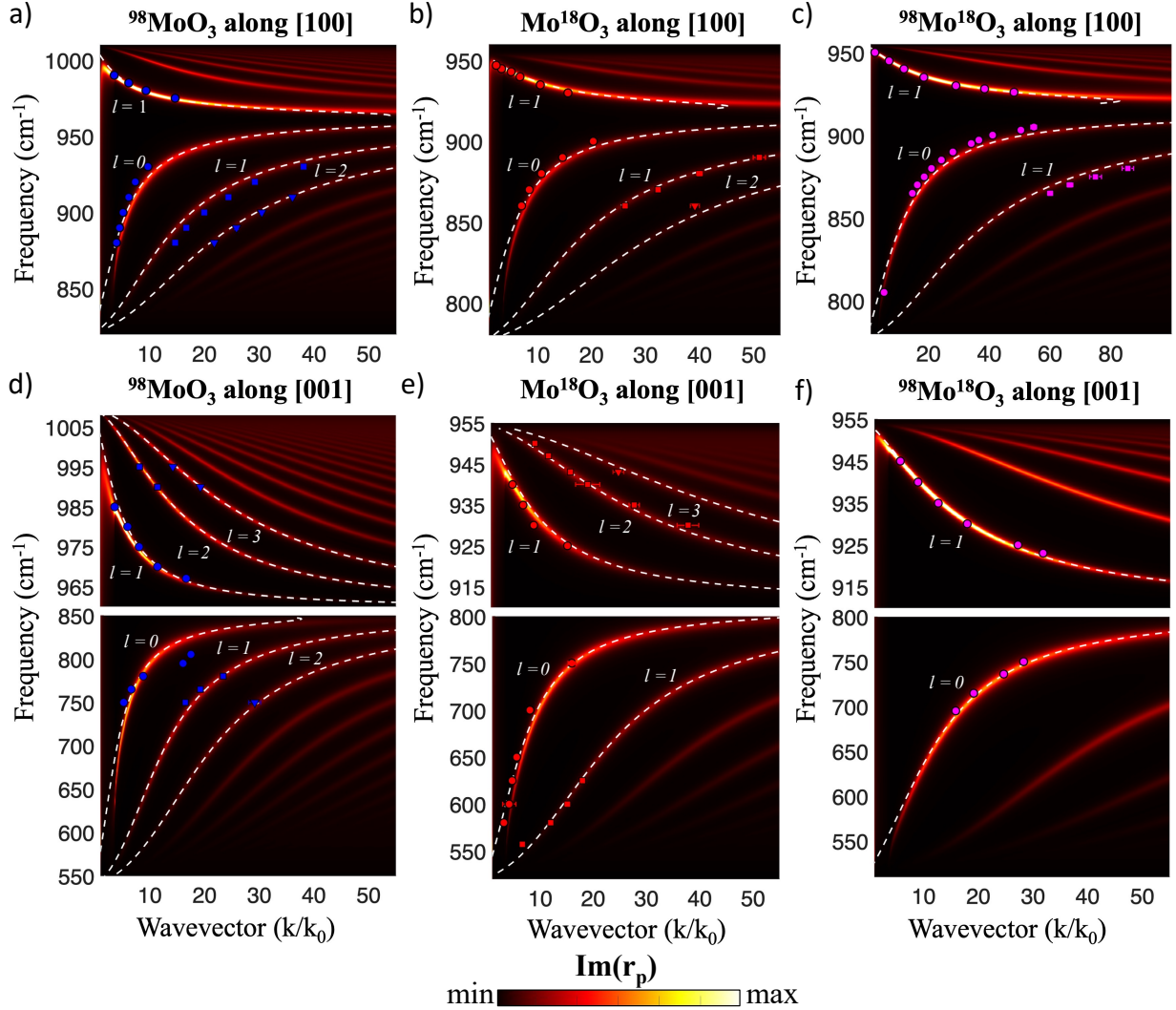

**Figure 3.** Experimental mapping the HPhP dispersion in different isotopically enriched α-MoO$_3$. **a-c)** TMM calculations of p-polarized imaginary-part of the reflection coefficient (Im(r$_p$)) along the [100] for $^{98}$MoO$_3$, Mo$^{18}$O$_3$, and $^{98}$Mo$^{18}$O$_3$, respectively. **d-f)** Similar TMM calculations and experimental data along the [001] for the same samples. The graphs were truncated in frequency gap between the RB$_1$ and RB$_3$ and rescaled so that the RB$_3$ is more visible to the reader. The analytical dispersion from Equation 3 is calculated for each isotopic variation and plotted as a white dashed line for all the experimentally observed orders. Data points extracted from s-SNOM images for each sample are overlayed in their respective plots. The error bars are propagated from the FFT analysis and laser linewidths.

To investigate the optical loss across our isotopically enriched samples, we report the polariton Q-factors for in all three RBs- defined by $Q = \frac{2\pi \text{Re}(k)}{\text{Im}(k)}$. From the complex polariton wavevector extracted from the FFT analysis, we calculate the Q-factors as a function of the thickness normalized wavevector ($k * d$) across all isotopic samples for each RB$_{1-3}$ in **Figure 4a-c,** respectively, (larger sample thickness HPhP Q-factors provided in **Figure S11.1**). This thickness normalized wavevector is easily obtained from manipulating **Equation 1** and discarding the free-space normalization. It is necessary to implement this unitless independent variable to compare HPhP Q-factors across samples varying in thickness, as the momentum $k$ scales inversely with the thickness $d$. This methodology has previously been employed by *Schultz et al.* to compare dispersions of different $^x$Mo isotopes in α-MoO$_3$.[46] Within the RB$_1$, we observe similar magnitudes of Q independent of which isotope was enriched (**Figure 4a**). Within this spectral region, we also measured naturally abundant α-MoO$_3$ and report its HPhP Q-factors (**Figure S11.2**). Importantly, however, using an FEL as an excitation source can lead to inconsistent Q-factors across isotopic samples due to the variable spectral bandwidth and signal-to-noise ratio (SNR) over a large frequency range such as the excitation frequencies chosen for the RB$_1$ (550-820 cm$^{-1}$). Likewise, detectors at lower frequencies operate with a lower sensitivity. Therefore, HPhPs of similar momenta will have different experimental excitation conditions when measured at different frequencies, which is unavoidable due to the red shift from oxygen enrichment. Aside from varying experimental conditions, recall from **Figure 1c** that the RB$_1$ TO is still significantly more damped than the other two corresponding TO phonons in RB$_2$ and RB$_3$. While we claim narrower linewidths from oxygen enrichment in the far-field and believe similar enhancements would be present in the near-field, our current measurements do not validate such an assumption. On the other hand, we observe a ~50% increase in Q-factor maxima within the RB$_2$ for both Mo$^{18}$O$_3$ and $^{98}$Mo$^{18}$O$_3$ with respect to $^{98}$MoO$_3$ and a slight shift in normalized wavevector where the Q-factor is maximized (**Figure. 4b**). Here, the Q-factor maxima increase is due to the moderate increase in phonon lifetimes from $^{18}$O enrichment, as previously discussed. We also lend this Q-factor enhancement to a slight increase in free space wavevector compression. The confinement factor ($\frac{k}{k_0}$) for the thickness-normalized wavevector is equal in both Mo$^{18}$O$_3$ and $^{98}$Mo$^{18}$O$_3$ yet greater than in $^{98}$MoO$_3$ (**Figure S12.1**). This is a result of the significant red shift due to $^{18}$O enrichment, where the HPhPs carry the same wavevector for a reduced free-space wavevector. The Q-factor maximizes at the wavevector where both the TO absorption and HPhP scattering rates are minimized - lower wavevectors are dominated by absorption loss ($\gamma_\alpha$) and higher wavevectors are dominated by scattering losses. *Knighton et al.* discusses the relation between phonon polariton dispersion and damping rate, where they show increasing polariton damping rates ($\gamma_p$) for increasing polariton wavevectors.[61] As a result, the polariton damping rate increases with wavevector as the HPhP branch approaches the asymptotic limit. Therefore, the increased scattering from the Mo isotopic disorder ($\gamma_{Mo}$) shares the same relationship as the damping rates due to the negligible change in Q-factor at low wavevector and the ~50% increase in Q across the higher wavevectors from Mo$^{18}$O$_3$ to $^{98}$Mo$^{18}$O$_3$. As a result of dual isotopic enrichment, HPhPs in $^{98}$Mo$^{18}$O$_3$ exhibit a higher Q-factor across a broader range of momenta in the RB$_2$.

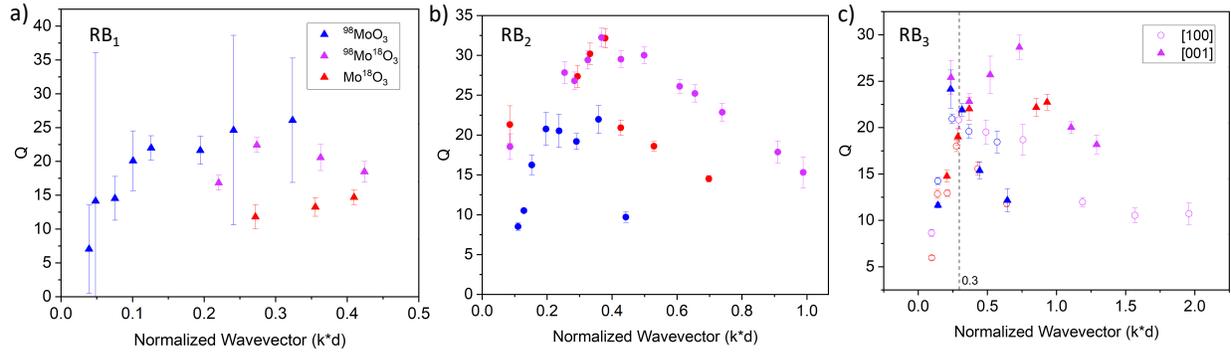

**Figure 4.** Isotopic α-MoO$_3$ polariton $Q$-factors across the MIR Reststrahlen bands. **a-c)** Polariton $Q$-factors in the RB$_1$, RB$_2$, and RB$_3$, respectively. All flakes measured for the results shown in Figures a and b are comparable in thickness (~200 nm). In fig 4c, $^{98}$MoO$_3$ and Mo$^{18}$O$_3$ are 460 nm and 440 nm thick while $^{98}$Mo$^{18}$O$_3$ is 200 nm. The circular data points depict the $Q$-factors extracted along the [100], while the triangular data points show $Q$ extracted along the [001] direction. The vertical dashed line in fig.4c is plotted as a visual aid to indicate the wavevector vicinity where the RB$_3$ anisotropic $Q$-factor splitting occurs.

Lastly, we expect a high sensitivity to the change of $Q$-factors along both directions of the RB$_3$, due to the ultra-narrow linewidth of the respective TO phonon. The elliptic behavior of the RB$_3$ is demonstrated through the distinct $Q$-factors of HPhPs propagating along the [100] or [001] (**Figure 4c** and **Figure S13.1** for more clarity). *Zhao et al* reports similar $Q$-factors for both $^{92}$MoO$_3$ and $^{100}$MoO$_3$ along the [100][45] compared to the $^{98}$MoO$_3$ $Q$-factors we report in this work. These findings corroborate the understanding that any Mo enrichment reduces the scattering loss due to isotopic disorder while the choice of isotope has minimal influence on the $Q$-factor maxima due to the small mass variance. In the large momenta regime ($k*d > 0.3$), we observe higher $Q$-factors in $^{98}$MoO$_3$ along the [100] than the [001]. This anisotropic relation of $Q$-factors in the RB$_3$ is due proximity of the RB$_3$ TO phonon (961 cm$^{-1}$) to the frequency where the $l$=0 branch approaches the asymptotic limit; ~961 cm$^{-1}$ along the [001] and ~971 cm$^{-1}$ along the [100]. Therefore, the differences in HPhP dispersion impose a larger $\gamma_\alpha$ on HPhPs along the [001] than the [100] - leading to higher $Q$-factors along the [100]. Surprisingly, we observe a stronger and inverted anisotropic response from $^{18}$O enrichment, where $Q$-factors are now greater along the [001] than along the [100]. The larger $Q$-factors along the [001] than along the [100] implies that $\gamma_\alpha$ is now considerably less than $\gamma_p$ - causing the in-plane axis of higher $Q$ HPhPs to flip from [100] to [001]. This observation further supports the modest RB$_3$ TO lifetime improvements that we expect from $^{18}$O enrichment. Lastly, we correlate these improvements to the ~100% enhancement of the $Q$-factor maxima along the [001] in Mo$^{18}$O$_3$ with respect to $^{98}$MoO$_3$. We do not compare the $Q$-factors in $^{98}$Mo$^{18}$O$_3$ to the other two isotopes due to a large discrepancy in sample thickness that the unitless wavevector cannot account for. However, we still expect similar enhancements in $Q$-factor maxima from $^{18}$O enrichment and the broadest wavevector range of high-$Q$ HPhPs to be supported in thin flakes of $^{98}$Mo$^{18}$O$_3$.

**Conclusion**

In this work, we investigated both avenues of optical and acoustic phonon atomic dependencies through far-field and near-field optical measurements and thermal conductivity measurements in isotopically enriched α-MoO$_3$. We provide an in-depth theoretical analysis that both validates and predicts various phonon scattering mechanisms present in α-MoO$_3$. Raman and FTIR spectroscopy were performed to initially observe the expected ~40 cm$^{-1}$ redshift from $^{18}$O enrichment and linewidth reductions. Consistent with this observation, our mode-resolved three-phonon scattering analysis confirms that the lifetime enhancement is not captured by a simple mass-scaling of scattering rates. Rather, it reveals that the $^{18}$O enrichment renormalizes the harmonic phonon spectrum and thereby reshapes the available three-phonon phase-space and coupling landscape, particularly those driven by oxygen oscillations, such that the net anharmonic linewidths decrease. We report the approximate dielectric functions of $^{98}$MoO$_3$, Mo$^{18}$O$_3$ and $^{98}$Mo$^{18}$O$_3$ from their respective phonon shifts. TDTR measurements were conducted across the same isotopes and validated with DFPT calculations. Further theoretical thermal analysis reveals the possibility to deterministically engineer directional thermal transport in isotopic α-MoO$_3$ flakes. The near-field IR properties were probed with s-SNOM to measure the HPhPs in each of the three MIR RBs. We experimentally report propagating higher order HPhP modes up to $l$=2 without imposing additional scatterers. The highly anisotropic character of each TO phonon from their respective RBs is reflected into the degree of HPhP $Q$-factor enhancements that are only observed due to isotopic enrichment. Through these efforts, we discuss the interchange of loss mechanisms at play in α-MoO$_3$ with respect to HPhP wavevector: predominantly being absorption and polariton damping rates. This work experimentally explores the intrinsic limit of tuning $Q$-factors in α-MoO$_3$ through isotopic enrichment. We demonstrate the benefits from a dual-isotopic $^{x}$Mo$^{18}$O$_3$ slabs in supporting enhanced HPhP $Q$-factors and higher wavelength compression than with $^{16}$O. Such a platform can have significant impact in applications such as isotopic heterostructure twist-optics, hyper-resolution imaging, and dispersion engineering.

**Methods**

*Sample growth and preparation:*

The α-MoO$_3$ crystals were primarily grown using a reactive vapor transport method which is detailed in a companion report[62]. In short, a molybdenum (naturally abundant or isotope-enriched) metal source is heated in an inert environment within a multi-zone tube furnace, where it is rapidly oxidized and vaporized by $O_2$ flow of select isotope enrichment. Numerous mm- to cm-scale α-MoO₃ crystals were grown after the precursor vapor condenses on the walls of the tube within the cooler furnace zone. The $^{98}$MoO₃ samples were grown using a similar non-reactive vapor transport method using $^{98}$Mo-enriched MoO₃ powder (IsoFlex USA, 98.42% enrichment) as the evaporating source and $N_2$ carrier gas. The free-standing crystals were then exfoliated with tape and transferred to silicon substrates until the desirable surface area and thicknesses were achieved for both far and near-field measurements.

*Density functional perturbation theory calculations:*

First-principles density functional perturbation theory (DFPT) calculations were performed using Quantum ESPRESSO[63] to compute interatomic force constants (IFCs) and three-phonon scattering rates for layered α-MoO₃. The van der Waals density functional vdW-DF-optB88 (vdw-df-obk8)[64] was used to accurately capture the weak interlayer interaction following previous validation for MoO₃.[65] Optimized norm-conserving Vanderbilt (ONCV) pseudopotentials were used for Mo and O, with a plane-wave kinetic energy cutoff of 120 Ry.[66] Self-consistent field calculations were converged to a threshold of $10^{-10}$ Ry.

Harmonic (second-order) IFCs were calculated using the finite displacement method in supercells commensurate with the desired phonon wavevector mesh. For the 3×3×1 supercell (144 atoms), a Γ-centered 4×4×3 k-point mesh was employed with 16 symmetry-inequivalent displacement configurations. The 3×3×2 supercell (288 atoms) used a 4×4×2 k-point mesh with 24 configurations. These IFCs were used to construct dynamical matrices and compute phonon dispersions.

Anharmonic (third-order) IFCs, which govern three-phonon scattering processes, were computed using a 2×2×1 supercell (64 atoms) with a denser 6×6×3 k-point mesh. A total of 6,680 symmetry-inequivalent configurations were required to fully sample the third-order force constant tensor. Non-analytic corrections (NAC) to the dynamical matrix, arising from long-range dipole–dipole interactions, were included using Born effective charge tensors and the high-frequency dielectric tensor obtained from density-functional perturbation theory as implemented in Quantum ESPRESSO *ph.x*. The NAC was evaluated along the principal crystallographic directions [100], [010], and [001], enabling resolution of the polarization-dependent LO–TO splitting for infrared-active modes at the Brillouin zone center.

*First-principles three-phonon scattering and linewidth calculations:*

Phonon dispersions, three-phonon scattering rates, and the lattice thermal conductivity were computed using the phono3py package[67], with the harmonic and anharmonic IFCs from the finite-displacement DFPT calculations as input.

Non-analytic corrections to the dynamical matrices were included in phono3py using the Born effective charges and dielectric tensor, following the Gonze and Lee formalism.[68] This treatment captures LO–TO splitting and polar-optical phonon behavior. The resulting phonon frequencies and eigenvectors, including NAC, were then used to evaluate three-phonon scattering processes.

Anharmonic phonon linewidths arising from three-phonon scattering were computed within the single-mode relaxation time approximation (SMRT) as implemented in phono3py, by evaluating the imaginary part of the phonon self-energy from all energy- and momentum-conserving three-phonon processes. Scattering rates were obtained from the third-order IFCs and phonon eigenvectors following the formalism of Togo, Chaput, and Tanaka.[67] Γ-point phonon linewidths were converged using a 36×36×12 Γ-centered q-point mesh at 300 K. Systematic tests of convergence verified this mesh density to be sufficient for the modes of interest up to the accuracy desired.

Isotope disorder scattering was included using the Tamura model.[69] The phonon–isotope scattering rate depends on the mass variance parameter:

$$g = \sum_i f_i \left(\frac{m_i}{\bar{m}} - 1\right)^2$$

where $f_i$ and $m_i$ are the fractional abundance and mass of isotope i, and $\bar{m}$ is the average atomic mass. For natural-abundance α-MoO₃, $g$ was evaluated including all stable Mo isotopes ($^{92}$Mo–$^{100}$Mo) and O isotopes ($^{16}$O, $^{17}$O, $^{18}$O). For the isotopically enriched samples, $g$ was recomputed assuming 98% $^{98}$Mo and 75-100% 18O enrichment, consistent with the experimental isotopic purities. In addition to modifying the mass variance parameter, the atomic masses used in constructing the dynamical matrix were adjusted to reflect the isotopic composition of each sample. Since the interatomic force constants are mass-independent, the same IFCs were used for all isotopic compositions, but the phonon frequencies and eigenvectors were recalculated using the isotope-averaged atomic masses. This captures the intrinsic frequency shifts and modified group velocities arising from isotopic mass substitution, which are distinct from the isotope-scattering effects encoded in $g$. The total scattering rate for each mode was taken as the sum of the anharmonic three-phonon and isotope-disorder contributions, $\Gamma_{total} = \Gamma_{anharmonic} + \Gamma_{iso}$, and the phonon lifetime was defined as $\tau = (2\,\Gamma_{total})^{-1}$.

The lattice thermal conductivity tensor κ was calculated at 300 K within the relaxation time approximation (RTA) by solving the linearized phonon Boltzmann transport equation using phono3py. A 36×36×12 Γ-centered q-mesh was used to sample the Brillouin zone. In this framework, $\kappa_{\alpha\beta}$ is calculated as:

$$\kappa_{\alpha\beta} = \left(\frac{1}{VN_q}\right) \Sigma_{q,s}\, C_{qs}\, v_{qs,\alpha} \otimes v_{qs,\beta}\, \tau_{qs}$$

where V is the crystal volume and $N_q$ is the number of q-points, and the sum runs over wavevectors q and phonon branches s. The mode heat capacity $C_{qs}$ and group velocity components $v_{qs,\alpha}$ are computed from the harmonic IFCs, while $\tau_{qs}$ is the phonon lifetime determined by the total scattering rate.

*Raman spectroscopy:*

Raman spectroscopy was performed using a LabRAM HR Raman microscope (Horiba) equipped with a 532 nm excitation laser and a 1800 gr/mm grating. Because of the polarization-dependent biaxial anisotropy of α-MoO$_3$, a consistent azimuthal orientation is required to achieve reproducible Raman peak intensity ratios.[70] Therefore, the α-MoO$_3$ flakes were supported on *c*-axis-oriented sapphire substrates and the [001] direction oriented parallel to the horizon of the instrument microscope. The spectra were normalized to the intensity of the strongest α-MoO$_3$ peak (819 cm$^{-1}$ for unenriched crystals) and the *x*-axis calibrated using the Rayleigh peak at 0 cm$^{-1}$. The instrument resolution, measured as the full width at half maximum (FWHM) of the Rayleigh scattering peak, was 0.8 cm$^{-1}$. The Raman linewidths were collected by averaging five measurements fitted with Lorentzian peak shapes.

*FTIR spectroscopy:*

The FTIR reflection measurements were performed through a Bruker Hyperion 2000 microscope equipped with the Bruker 15x Grazing Angle Objective (GAO) coupled to a Bruker Vertex70v FTIR spectrometer and collected with a broadband MCT detector (400 cm$^{-1}$ - 8000 cm$^{-1}$). Both a ZnSe beam splitter and Polyethylene linear polarizer (~600 cm$^{-1}$ - 5000 cm$^{-1}$) were used to obtain polarized reflection spectra of isotopically enriched α-MoO$_3$. The 15x GAO was employed for its viewing mode with near normal incidence of ~4°, where s and p polarization states nearly independently stimulate each in-plane direction. When the 15x GAO is switched to its grazing mode (~86°), we maintain a p-polarization state and instead, rotate the crystal to excite each in-plane direction. Internal microscope apertures of 50 x 50 μm were used to ensure light is collected from the crystal. All spectra were collected with a 2 cm$^{-1}$ spectral resolution and normalized with the spectra from a gold mirror taken with the same experimental conditions.

*Near-field characterization:*

The study of highly confined polaritons requires optical measurements with subdiffractional resolution, which can be achieved via s-SNOM measurements.[71] In s-SNOM, light illuminates and is scattered by a metal coated AFM tip; the scattered light is subsequentially detected by a suitable optical detector. The AFM tip acts as an antenna, allowing for simultaneous access and scattering of the near-field from the sample directly below the tip, with a spatial resolution corresponding to the tip radius (typically ~20 nm).[72,73] Additionally, the highly confined field at the tip apex provides enough momentum to couple to HPhPs.[74] The tip simultaneously scatters the near-field from propagating polaritons that were either tip launched and edge reflected or directly edge launched. For the optical near-field measurements presented in this publication, two commercial s-SNOM devices by neaspec (Attocube GmbH) were utilized, coupled to two different light sources. Measurements in $RB_2$ and $RB_3$ were performed using a tunable quantum cascade laser (QCL; MIRCAT by Daylight Solutions). For measurements using this light source, an interferometric setup with an oscillating mirror in the reference arm (pseudo-heterodyne detection scheme) was employed, allowing for a suppression of the far-field and the separation of the optical amplitude and phase.[75] Measurements in $RB_1$ were instead performed utilizing the tunable free-electron laser FELBE located at the Helmholtz-Zentrum Dresden-Rossendorf (HZDR), which allows for accessing the required THz frequencies.[17,54,55,76] The corresponding setup uses the self-homodyne detection scheme instead due to the challenges in achieving pseudo-heterodyne detection using a FEL. As the wavelength of the far-field is roughly an order of magnitude longer than that of the measured polaritons, a separation of the two phenomena is possible in post-processing.[17]

**Acknowledgements**

T.S.A. acknowledges support from the National Science Foundation Graduate Research Fellowship Program under Grant No. 2444112. R.W.S. and J.P.M. acknowledges support from the Army Research Office under Grant No. W911NF-21-1-0119. D.H. and P.E.H appreciate support from the Office of Naval Research under Grant No. N00014-23-1-2630. J.D.C., J.D.G., J.B.H., M.O., J.E.B., Y.K., M.H., A.C.B., R.B., N.S., T.G.F., and F.J. acknowledge support from Multi-University Research Initiative (MURI) on Twist-Optics, sponsored by the Office of Naval Research under Grant No. N00014-23-1-2567. M.L. acknowledges support from National Science Foundation under Grant No. 2406022. C.R. acknowledges support from Office of Naval Research under Grant N00014-24-1-2284 – Materials Design of Optical Strong Coupling for Advanced Imaging and Sensing Applications through the University of Minnesota. A.S. acknowledges support from the Department of Energy, Basic Energy Sciences (grant DE-FG02-09ER4655). R.N. acknowledges support from Army Research Office under Grant No.W911NF-24-2-0195. U.A.L. acknowledges support from the HOTMETA project under the PRIN 2022 MUR program funded by the European Union − Next Generation EU − PNRR − M4C2, investimento 1.1 − "Fondo PRIN 2022" − HOT-carrier METasurfaces for Advanced photonics (HOTMETA), contract no. 2022LENW33 − CUP: D53D2300229 0006. The calculations of the DFPT forces and phonon lifetimes of $MoO_3$ used computational resources from the Texas Advanced Computing Center (TACC) at The University of Texas at Austin, funded by the National Science Foundation (NSF) award 1818253, through allocation DMR21077. Parts of this research were carried out at the FELBE Center for High-Power Radiation Sources at the Helmholtz- Zentrum Dresden-Rossendorf e.V., a member of the Helmholtz Association. G.C., A.C.B., M.W., and A.P. acknowledge support by the Max Planck Society. M.O., F.G.K., J.W., J.M.K., L.M.E., and S.C.K. acknowledge the financial support by the Bundesministerium für Bildung und Forschung (BMBF, Federal Ministry of Education and Research, Germany, Project Grant 05K19ODB, and 05K22ODA) and by the Deutsche Forschungsgemeinschaft (DFG, German Research Foundation) through the project CRC1415 as well as under Germany's Excellence Strategy through Würzburg-Dresden Cluster of Excellence on Complexity and Topology in Quantum Matter—ct.qmat (EXC 2147, project-id 390858490).

**Conflict of Interest**

The authors declare no conflict of interest.

**Author Contributions**

R.W.S., J.P.M., and J.D.C. conceived the idea. R.W.S. grew the bulk isotopically enriched α-$MoO_3$ crystals under supervision of J.P.M. J.D.G. and J.B.H. executed all the DFPT calculations and analysis under supervision of F.J. R.W.S. performed all the Raman measurements and fittings supervised by J.P.M. T.S.A. performed all the sample preparations for FTIR and SNOM measurements. T.S.A performed the FTIR reflection measurements supervised by J.D.C. T.S.A. performed the majority of the SNOM measurements for the $RB_2$ and $RB_3$ across all isotopes and M.O. performed the $RB_3$ measurements for the thicker $Mo^{18}O_3$ sample; supervised by J.D.C. M.O., G.A.P., M.L., F.G.K., J.W., C.R., J.E.B., Y.K., A.S.S., R.N., G.C., U.A.L., A.C.B., R.B., N.S., N.C.B., J.M.K., M.W., T.G.F., and A.P. carried out the SNOM measurements at the Helmholtz-Zentrum Dresden-Rossendorf (HZDR) with the tunable free-electron laser FELBE supervised by M.W., L.M.E., and S.C.K. T.SA. and M.O. performed the HPhP momenta analysis from all the SNOM measurements supervised by J.D.C. T.S.A., R.W.S., J.D.G., J.B.H., and D.H. prepared the figures and drafted the manuscript with thorough feedback and revisions from M.O., G.A.P., M.L., M.H., U.A.L., R.B., S.C.K., T.G.F., P.E.H., F.J., J.P.M., and J.D.C.

**Data Availability Statement**

**Supplementary material: Tailoring phonon-driven responses in α-MoO$_3$ through isotopic enrichment**

Thiago S. Arnaud[1,2], Ryan W. Spangler[3], Johnathan D. Georgaras[4], Jonah B. Haber[4], Daniel Hirt[5], Maximilian Obst[1,2,6,7], Gonzalo Álvarez-Pérez[8,9], Mackey Long III[1,2], Felix G. Kaps[6,7], Jakob Wetzel[6,7], Courtney Ragle[1,2], John E. Buchner[1,2], Youngji Kim[2], Aditha S. Senarath[1,2], Richarda Niemann[2], Mingze He[2], Giulia Carini[9], Unai Arregui-Leon[9,10], Akash C. Behera[9], Ramachandra Bangari[12], Nihar Sahoo[12], Niels C. Brumby[9], J. Michael Klopf[11], Martin Wolf[9], Lukas M. Eng[6,7], Susanne C. Kehr[6,7], Thomas G. Folland[12], Alexander Paarmann[9], Patrick E. Hopkins[5,13,14], Felipe Jornada[4]*, Jon-Paul Maria[3]*, Joshua D. Caldwell[1,2]*

[1] Interdisciplinary Material Science, Vanderbilt University, Nashville 37235, TN, USA

[2] Department of Mechanical Engineering, Vanderbilt University, Nashville 37235, TN, USA

[3] Department of Materials Science and Engineering, The Pennsylvania State University, University Park 16802, PA, USA

[4] Department of Materials Science and Engineering, Stanford University, Stanford, CA 94305, USA

[5] Department of Mechanical and Aerospace Engineering, University of Virginia, Charlottesville, Virginia 22904, USA

[6] Institute of Applied Physics, TUD Dresden University of Technology, Dresden 01187, Germany

[7] Würzburg-Dresden Cluster of Excellence - EXC 2147 (ct.qmat), Dresden 01062, Germany

[8] Istituto Italiano di Tecnologia, Center for Biomolecular Nanotechnologies, Lecce, Italy

[9] Department of Physcial Chemistry, Fritz Haber Institute of the Max Planck Society, Faradayweg 4-6, 14195, Berlin, Germany

[10] Department of Physics, Politecnico di Milano, Piazza Leonardo da Vinci 32, 20133, Milan, Italy

[11] Institute of Radiation Physics, Helmholtz-Zentrum Dresden-Rossendorf, Dresden 01328, Germany

[12] Department of Physics and Astronomy, University of Iowa, Iowa City 52242, IA, USA

[13] Department of Materials Science and Engineering, *University of Virginia*, Charlottesville, Virginia 22904, USA

[14] Department of Physics, *University of Virginia*, Charlottesville, Virginia 22904, USA

* Correspondence to: josh.caldwell@vanderbilt.edu, jpm133@psu.edu, and jornada@stanford.edu

**Table of Contents**

# Section S1: *Ab Initio* optical and acoustic phonon properties

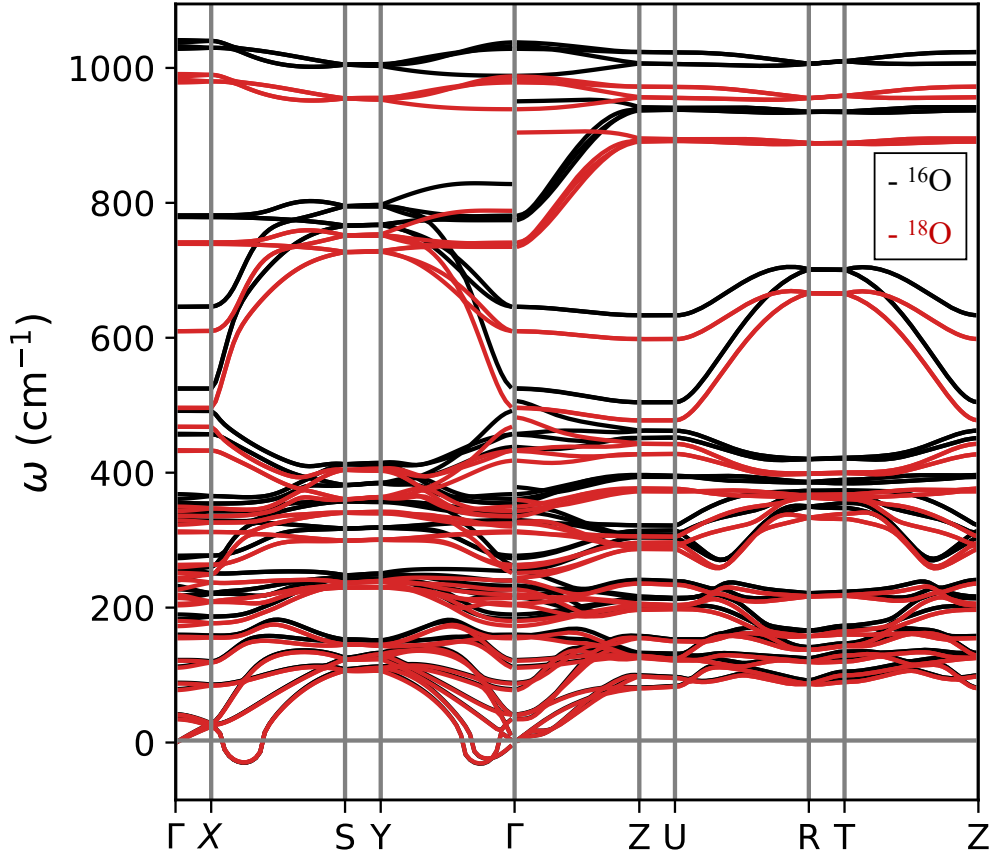

**Figure S1.1:** First-principles calculations of the phonon dispersion in α-MoO₃ (black) and α-Mo¹⁸O₃ (red).

We first compute the first-principles harmonic phonon dispersion of α-MoO₃ for all isotope compositions by density-functional perturbation theory (implemented in Quantum Espresso). Crucially, we include non-analytic corrections (NAC) to the dynamical matrix to account for the macroscopic electric fields generated by polar vibrations. This correction captures the longitudinal optical–transverse optical (LO–TO) splitting at the -point which opens the Reststrahlen bands (RB$_{1-3}$) essential to the hyperbolic optical behavior discussed in this work. **Figure S1.1** compares the full dispersion for naturally abundant α-MoO₃ (¹⁶O, black) and fully ¹⁸O-enriched α-MoO₃ (red). As expected for a lattice in which oxygen is the lighter sublattice, oxygen substitution primarily red-shifts the mid- and high-frequency optical branches, including the phonons forming the three MIR Reststrahlen bands (RB$_{1-3}$), while leaving the acoustic branches and most low-frequency optical modes nearly unchanged. These trends are consistent with the red shifts of the Raman, IR, and HPhP resonances observed in the main text and set the baseline for the isotope-dependent Raman, IR, and thermal analyses presented in Sections S2, S4, and S6.

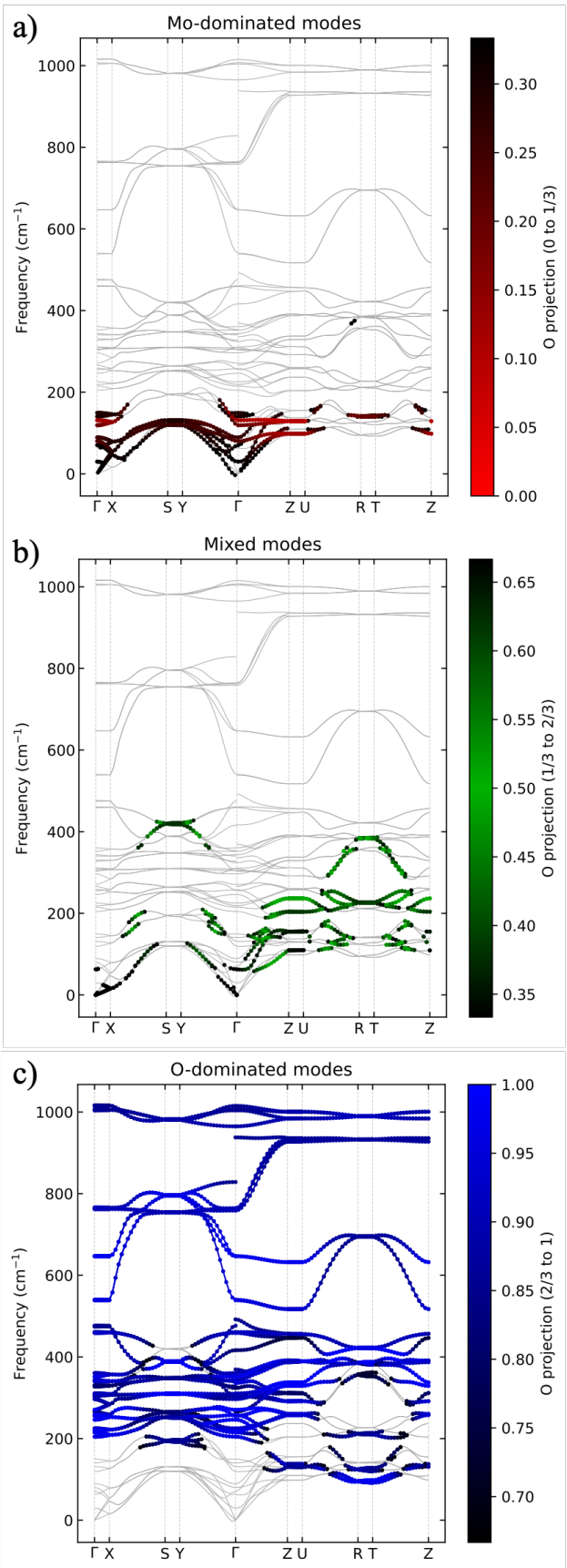

**Figure S1.2:** *Ab initio* atom-resolved phonon dispersion of α-MoO3. Phonon band structure along high-symmetry paths, decomposed by atomic character based on the squared eigenvector projections onto Mo and O sublattices. (Left) Mo-dominated modes, where the eigenvector projection onto oxygen atoms is less than 1/3, shown with a red color scale indicating Mo character (100% = pure Mo motion). (Center) Mixed-character modes with intermediate O projection (1/3 to 2/3), displayed with a green color scale where maximum intensity (100%) corresponds to equal Mo/O participation. (Right) O-dominated modes with O projection exceeding 2/3, shown in blue color scale indicating O character (100% = pure O motion). Grey lines show all phonon bands for reference. In Figure S1.4 and Figure S1.8 we use ~180 cm$^{-1}$ as the threshold between acoustic and optical modes for analysis.

To quantify how different atoms participate in each phonon, we project the eigenvectors onto Mo and O sublattices and classify the modes as Mo-dominated, mixed, or O-dominated (**Figure S1.2**). The modes below ~200 cm$^{-1}$ are predominantly Mo-like and correspond to acoustic and low-energy optical branches that control heat transport, whereas the mid- to high-frequency optical modes above ~500 cm$^{-1}$ are strongly O-dominated and underlie the IR-active Reststrahlen bands and Raman-active stretching modes. Because the heat-carrying acoustic modes are dominated by the heavier Mo atoms while the HPhP-driving optical modes are dominated by the lighter O atoms, α-MoO3 offers a unique platform to decouple thermal and optical engineering via site-specific isotope selection. For the IR-active TO modes corresponding to the MIR RBs, we additionally visualize the real-space atomic translations in the primitive cell (**Figure S1.3**). As expected of the lighter mass driving the optical phonons, the oxygen atoms are the predominant drivers of the transverse oscillations. Due to the orthorhombic crystal structure of α-MoO3, the TO phonons along each crystallographic direction is dictated by the number of Mo atoms the oxygen is bonded to. The lowest energy TO phonon along the [001] belonging to the RB$_1$ is dictated by the oxygen atom labeled as O(1), where it is bonded to three Mo atoms shown in **Figure S1.3a**. Subsequently, the RB$_2$ and RB$_3$ TO phonons are driven by oxygen atoms bonded to two and one Mo atom(s), respectively (**Figure S1.3b-d**). Taken together, these mode patterns and linewidth calculations suggest a qualitative correlation between the local constraint on the principal oxygen atom and the HPhP Q-factor: TO modes in which the relevant O site is more strongly bonded (and thus more restricted) exhibit smaller IR/Raman linewidths and, correspondingly, higher calculated Q-factors for the associated Reststrahlen band. In the following sections, we use first-principles calculations to confirm certain trends observed in experimental data. We calculate dispersions and mode characters to analyzing isotope-dependent Raman linewidths (Section S2), IR-active TO linewidths (Section S4), and lattice thermal conductivity and its decomposition into acoustic and optical contributions (Section S6).

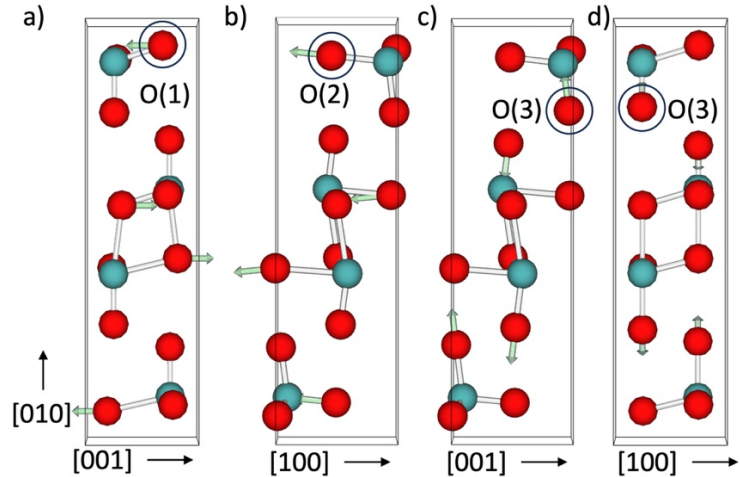

**Figure S1.3:** Visualized eigen displacements occurring for each TO phonon in a unit cell of α-MoO$_3$ corresponding to the RB$_1$ in a), RB$_2$ in b), and RB$_3$ in c) and d), respectively. The principal oxygen atom undergoing displacement is labeled in each of their respective figures.

## Section S2: *Ab Initio* Raman-active phonon modes and lifetimes

The Raman shifts and linewidths extracted from the experimental spectra in Figure 1a are compared with the DFPT calculated values between equivalent isotopes in **Table S2.1-4**. The Raman spectra are dominated by three high-frequency modes, labeled B$_{1g}$, A$_g$, and B$_{2g}$, which lie in the upper part of the optical manifold of the phonon dispersion (Figure S1.1). These modes sit well above the low-frequency Mo-dominated acoustic and optical branches and reside in the region where the eigenvectors are strongly oxygen-like, consistent with the large Raman intensities extracted from first-principles Raman calculations and from the experimental spectrum of natural α-MoO$_3$. Using the atom-resolved phonon dispersion (Figure S1.2), we classify the experimentally observed Raman-active modes as predominantly O-dominated and further distinguish whether their displacements are mainly in-plane or out-of-plane, which is the key descriptor for how they couple to in-plane versus out-of-plane optical fields.

**Table S2.1:** Comparison of Raman shifts and linewidths between experimental and *ab initio* values in naturally abundant α-$MoO_3$

| Peak | Exp. $\omega$ $[cm^{-1}]$ | Ab Initio $\omega$ $[cm^{-1}]$ | Exp. $\gamma$ $[cm^{-1}]$ | Ab initio $\gamma$ $[cm^{-1}]$ |
|---|---|---|---|---|
| 1 ($B_{1g}$) | 665 | 640 | 10.18 | 8.71 |
| 2 ($A_g$) | 820 | 762 | 7.65 | 12.37 |
| 3 ($B_{2g}$) | 995 | 1012 | 1.15 | 0.659 |

**Table S2.2:** Comparison of Raman shifts and linewidths between experimental and *ab initio* values in $^{98}MoO_3$

| Peak | Exp. $\omega$ $[cm^{-1}]$ | Ab Initio $\omega$ $[cm^{-1}]$ | Exp. $\gamma$ $[cm^{-1}]$ | Ab initio $\gamma$ $[cm^{-1}]$ |
|---|---|---|---|---|
| 1 ($B_{1g}$) | 665 | 640 | 9.97 | 8.58 |
| 2 ($A_g$) | 818 | 761 | 7.56 | 12.38 |
| 3 ($B_{2g}$) | 994 | 1011 | 0.77 | 0.665 |

**Table S2.3:** Comparison of Raman shifts and linewidths between experimental and *ab initio* values in $Mo^{18}O_3$

| Peak | Exp. $\omega$ $[cm^{-1}]$ | Ab Initio $\omega$ $[cm^{-1}]$ | Exp. $\gamma$ $[cm^{-1}]$ | Ab initio $\gamma$ $[cm^{-1}]$ |
|---|---|---|---|---|
| 1 ($B_{1g}$) | 630 | 610 | 9.57 | 7.66 |
| 2 ($A_g$) | 778 | 729 | 7.42 | 11.98 |
| 3 ($B_{2g}$) | 947 | 971 | 1.84 | 0.597 |

**Table S2.4:** Comparison of Raman shifts and linewidths between experimental and *ab initio* values in $^{98}Mo^{18}O_3$

| Peak | Exp. $\omega$ $[cm^{-1}]$ | Ab Initio $\omega$ $[cm^{-1}]$ | Exp. $\gamma$ $[cm^{-1}]$ | Ab initio $\gamma$ $[cm^{-1}]$ |
|---|---|---|---|---|
| 1 ($B_{1g}$) | 630 | 609 | 9.55 | 7.75 |
| 2 ($A_g$) | 777 | 728 | 7.39 | 11.9 |
| 3 ($B_{2g}$) | 945 | 969 | 1.2 | 0.601 |

The subset of modes analyzed here was selected using the Raman implementation in Quantum ESPRESSO, which evaluates the derivative of the macroscopic dielectric tensor with respect to each normal mode within DFPT and from this, constructs mode-resolved Raman tensors and intensities. Modes with the largest calculated Raman activity and frequencies matching the experimental peaks were then used for detailed linewidth analysis. Within the single-mode relaxation-time approximation, the intrinsic anharmonic linewidth of a Raman-active mode $\lambda$ at wavevector $\Gamma$ is obtained from the imaginary part of the phonon self-energy built from third-order interatomic force constants, which we evaluate using the standard three-phonon expression[1]:

$$\Gamma_\lambda(\omega) = \frac{18}{\hbar^2} \sum_{\lambda',\lambda''} |\Phi_{-\lambda\lambda'\lambda''}|^2 [(n_{\lambda'} + n_{\lambda''} + 1)\delta(\omega - \omega_{\lambda'} - \omega_{\lambda''}) \\ + (n_{\lambda'} - n_{\lambda''})(\delta(\omega + \omega_{\lambda'} - \omega_{\lambda''}) - \delta(\omega - \omega_{\lambda'} + \omega_{\lambda''}))]$$

[Eq.S2.1]

where $\lambda \equiv (\mathbf{q}, j)$ labels a mode, $\Phi_{-\lambda\lambda'\lambda''}$ are the three-phonon interaction matrix elements, $\omega_\lambda$ and $n_\lambda$ are the phonon frequency and Bose–Einstein occupation, and the three δ–functions describe decay and absorption/emission processes; the total scattering rate is $1/\tau_\lambda = 2\Gamma_\lambda$. This formalism follows standard treatments of third-order phonon–phonon scattering and is equivalent to the expressions implemented in phono3py and related BTE solvers.

Finally, the mode-resolved three-phonon scattering analysis in **Figure S2.1**, together with the summary statistics in **Tables S2.5** and **S2.6**, decomposes $\Gamma_\lambda$ for the three experimentally observed Raman modes in naturally abundant α-MoO$_3$ into contributions from acoustic–acoustic (A+A), acoustic–optical (A+O), optical–acoustic (O+A), and optical–optical (O+O) channels and into different regions of the Brillouin zone via the histogram of $|\mathbf{q}|/|\mathbf{q}_{BZ}|$. The corresponding maps of daughter-modes character — referring to the two modes other than the primary parent mode in the momentum- and energy-conserving scattering triplet — show that the linewidths of these modes are dominated by scattering into O-dominated daughters, with A+O and O+O processes providing the largest fraction of $\Gamma_\lambda$. The tabulated fractions make explicit that channels involving two O-like daughters (O–O region in Figure S2.1) account for most of the anharmonic broadening, consistent with the picture that $^{18}$O enrichment primarily modifies the phase-space and matrix elements of O-dominated decay pathways and thereby drives the systematic linewidth reductions discussed above and in the main text.

**Table S2.5:** Acoustic/optical decomposition of the three-phonon scattering channels for Raman-active modes.

| Scattering Channel | Mode $B_{1g}$ Linewidth (cm$^{-1}$) | Mode $A_g$ Linewidth (cm$^{-1}$) | Mode $B_{2g}$ Linewidth (cm$^{-1}$) |
|---|---|---|---|
| A+A | 0.00 | 0.000 | 0.000 |
| A+O | 1.46 | 2.49 | 0.274 |
| O+A | 1.46 | 2.49 | 0.274 |
| O+O | 5.79 | 7.38 | 0.111 |
| Total | 8.71 | 12.37 | 0.659 |

**Table S2.6:** Mo/O atomic-character decomposition of three-phonon scattering channels for Raman-active modes.

| Atom Projection (1-2) | Mode $B_{1g}$ Linewidth (cm$^{-1}$) | Mode $A_g$ Linewidth (cm$^{-1}$) | Mode $B_{2g}$ Linewidth (cm$^{-1}$) |
|---|---|---|---|
| Mo-Mo | 0.00 | 0.000 | 0.000 |
| Mo-Mixed | 0.38 | 0.48 | 0.015 |
| Mo-O | 0.83 | 1.43 | 0.193 |
| Mixed-Mixed | 0.28 | 0.56 | 0.012 |
| O-Mixed | 1.90 | 2.40 | 0.156 |
| O-O | 5.32 | 7.50 | 0.283 |
| Total | 8.71 | 12.37 | 0.659 |

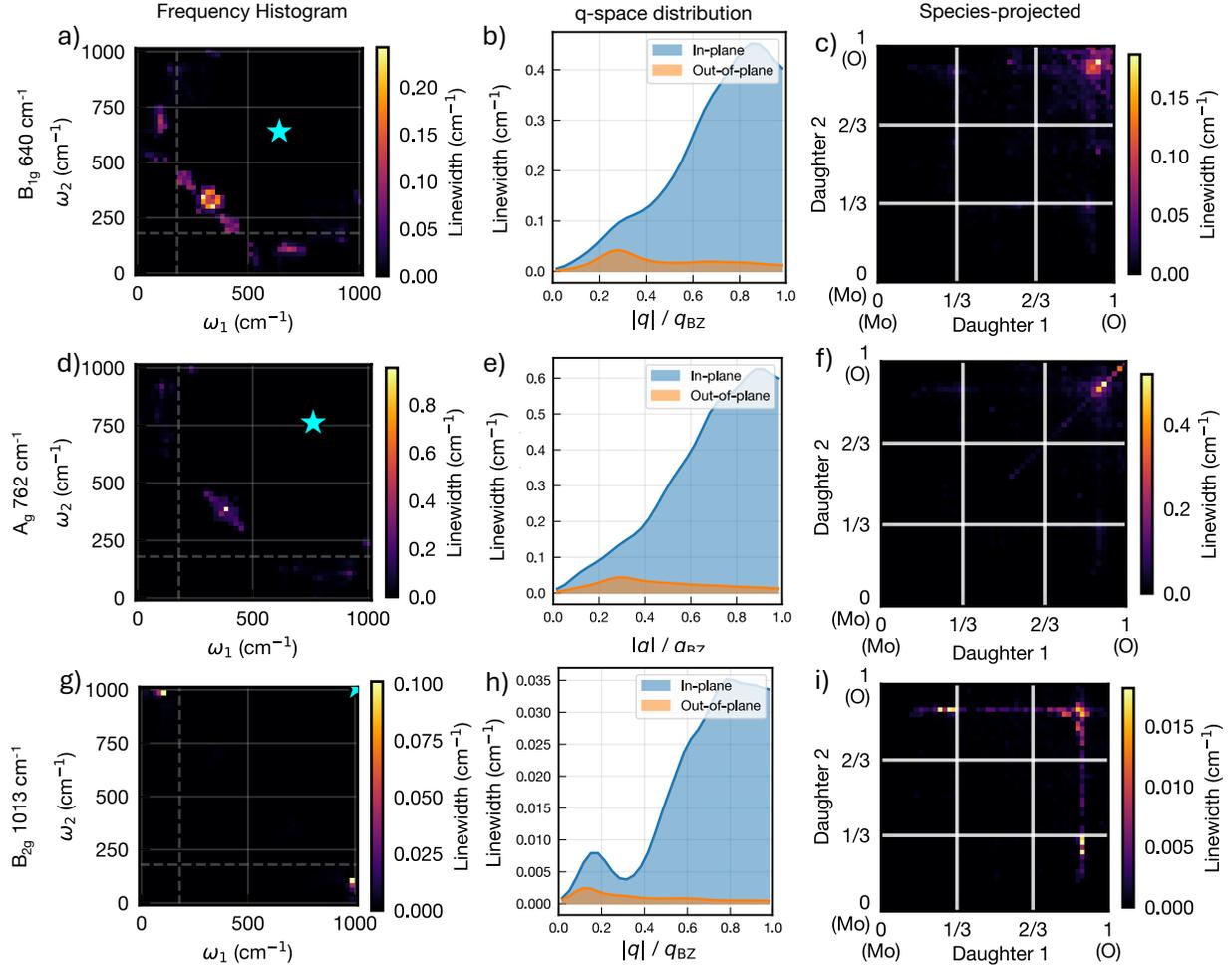

**Figure S2.1:** *Ab initio* mode-resolved three-phonon scattering analysis for Raman-active modes. Decomposition of the anharmonic phonon linewidth $\Gamma$ for three characteristic Raman-active modes at the $\Gamma$-point: mode $B_{1g}$ ($\omega_0 = 640\,\text{cm}^{-1}$, $\Gamma = 8.7\,\text{cm}^{-1}$), mode $A_g$ ($\omega_0 = 762\,\text{cm}^{-1}$, $\Gamma = 12.4\,\text{cm}^{-1}$), and mode $B_{2g}$ ($\omega_0 = 1013\,\text{cm}^{-1}$, $\Gamma = 0.66\,\text{cm}^{-1}$), calculated at $T = 300\,\text{K}$ using phono3py. (a,d,g) Daughter phonon frequency distributions ($\omega_1$ vs $\omega_2$) weighted by their contribution to the total linewidth. The dashed grey lines at $180\,\text{cm}^{-1}$ demarcate the acoustic–optical boundary. The cyan star marks the parent mode frequency. (b,e,h) Momentum-space distribution of scattering processes, showing the linewidth contribution as a function of normalized momentum transfer $|\mathbf{q}|/q_{BZ}$; Blue and orange curves denote in-plane ($q_\parallel$) and out-of-plane ($q_\perp$) components, respectively. (c,f,i) Daughter phonon character distributions showing the oxygen projection of daughter modes 1 and 2. Values near 0 correspond to predominantly Mo-centered vibrations, while values near 1 indicate O-dominated motion. The white grid lines at 1/3 and 2/3 delineate Mo-Mo, Mixed, and O-O character zones. Color intensity represents the scattering rate contribution written in linewidth notation (cm$^{-1}$) of the triplets of that character bin.

## Section S3: FTIR measurements of isotopic α-MoO₃ flakes

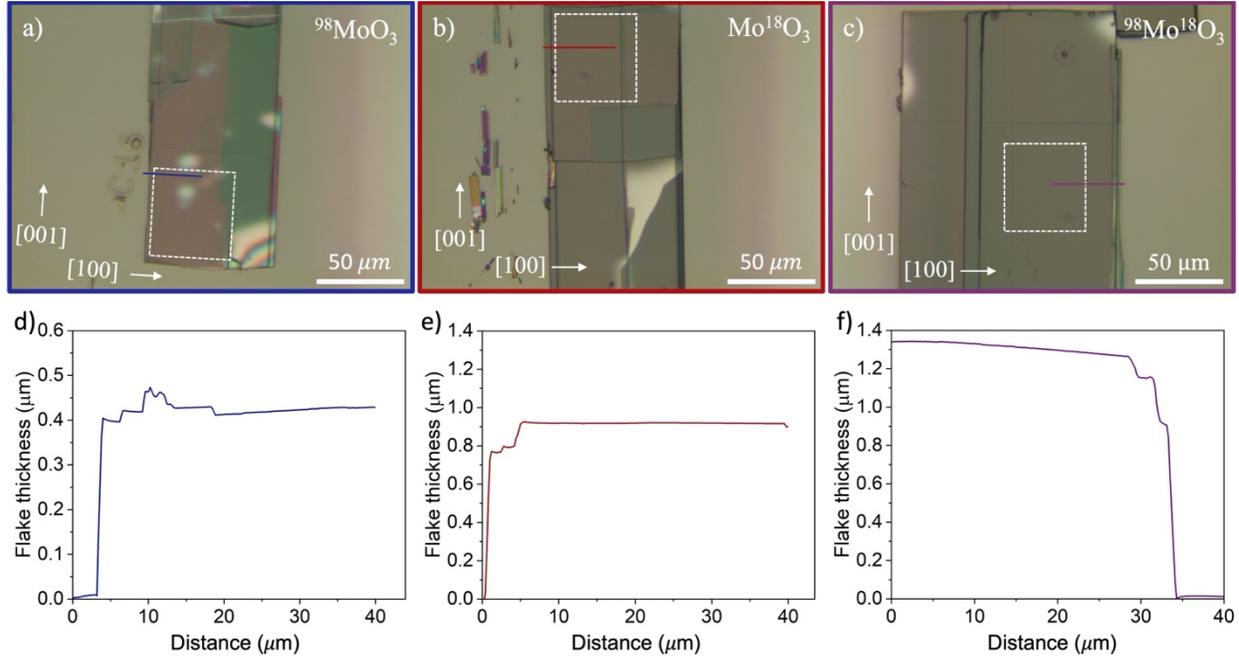

**Figure S3.1:** Visible microscope images of isotopic α-MoO₃ flakes (a-c) and their respective thicknesses extracted from AFM measurements (d-f). In a-c, the dashed white box insets are where FTIR reflection spectra was collected, and the colored bold lines are where the AFM measurement was taken.

Flakes of each isotopic variation were exfoliated with sufficiency surface area for FTIR measurements (**Figure S3.1**). The polarized FTIR reflection spectra are plotted with frequency markers that approximate the RB$_2$ TO phonon frequencies for the respective isotopic enrichments, shown in **Figure S3.2a**. We approximate the onset of the RB$_2$ as $828\,\text{cm}^{-1}$ for $^{98}\text{MoO}_3$ and $788\,\text{cm}^{-1}$ for $\text{Mo}^{18}\text{O}_3$ and $^{98}\text{Mo}^{18}\text{O}_3$. While these are crude approximations compared to fitting the reflection spectra with the TO LO model, we identify the expected phonon redshift from $^{18}\text{O}$ enrichment of $\sim 40\,\text{cm}^{-1}$ and validate the similar redshifts between Raman and IR active phonons. Shifting focus to the RB$_3$ expanded in **Figure S3.2b**, the redshifts are further validated in the RB$_3$ where TO phonon absorption (dotted black line) is present at $914\,\text{cm}^{-1}$ for both $\text{Mo}^{18}\text{O}_3$ and $^{98}\text{Mo}^{18}\text{O}_3$ but much less visible for $^{98}\text{MoO}_3$ at $966\,\text{cm}^{-1}$. While the LO phonon is not IR-active, its spectral position closely correlates with the upper frequency limit of the RB$_3$ where the HPhP branches converge. At this convergence, the DOS across wavevectors is significantly higher; resulting in a reflection peak. This correlative identification of the RB$_3$ LO phonons (dashed black line) are identified at $1000\,\text{cm}^{-1}$ for $^{98}\text{MoO}_3$ and $954\,\text{cm}^{-1}$ for $\text{Mo}^{18}\text{O}_3$ and $^{98}\text{Mo}^{18}\text{O}_3$. These TO LO phonon frequencies show good agreement with the reported value for naturally abundant α-MoO₃ at 956.7 cm⁻¹ and 1006.9 cm⁻¹, respectively[2]. Since this correlative LO peak is dependent on the HPhP DOS, we believe the proximity of the crystal edges and fractures serve as a scattering site to excite the HPhPs.

The out-of-plane phonons were excited by collecting reflection spectra at grazing incidence (**Figure S3.2c**), observed in the smaller ratio of reflectance between the RB₂ and the reflection peak near the LO phonon of the RB₃. Further investigation of this regime (**Figure S3.2d**) reveals a more accurate identification of the correlative LO reflection peak for ⁹⁸MoO₃ at 1004 cm⁻¹ with respect to the naturally abundant frequency at 1006.9 cm⁻¹ [2]. We also attribute the higher reflectivity from the correlative LO peak due to the grazing incidence illumination. At such a shallow angle, total internal reflection is achieved on the surface and creates evanescent fields which sufficient momenta to excite the HPhPs. Due to the nosier spectra from a grazing angle of incidence, the RB₃ TO phonons are more difficult to discern. Nevertheless, we show excellent agreement between the reported values in literature and the ~5% redshifted phonons. Furthermore, these values are in good agreement with the predicted phonon frequencies reported in Table 1 that were slightly tuned with near-field data.

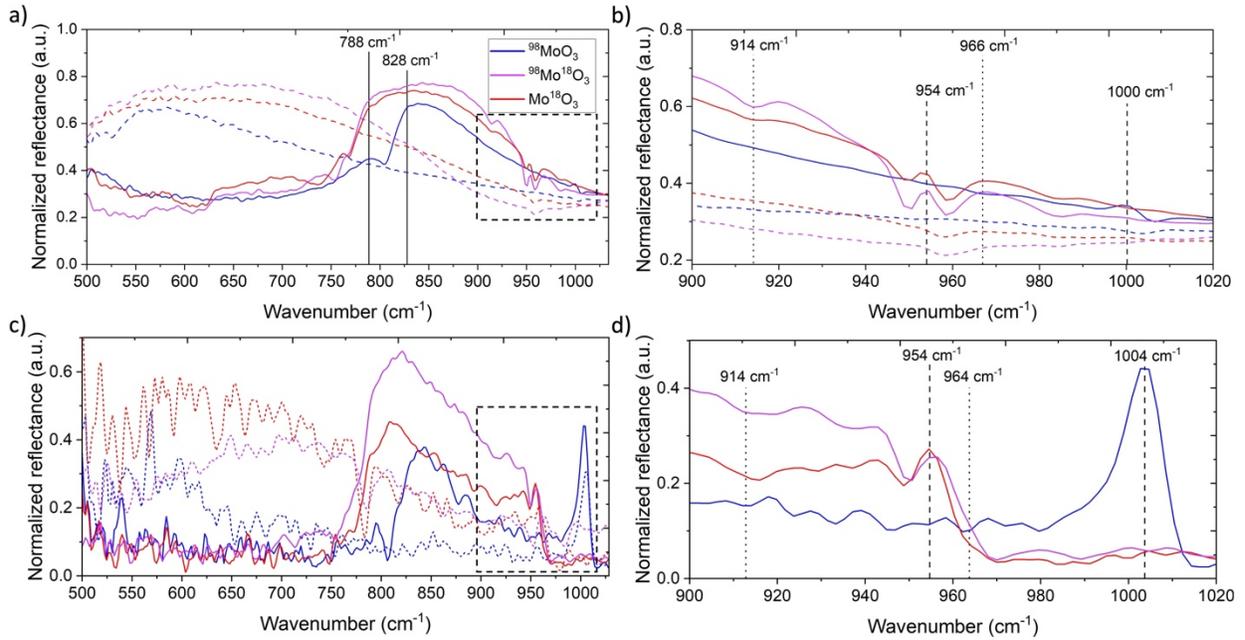

**Figure S3.2:** FTIR reflection spectra taken at near-normal incidence with the 15x Bruker GAO and the RB₂ TO phonons labeled (a) and an expanded view of phonons in the RB₃ (b). Reflection spectra taken at grazing incidence with the 15x Bruker GAO (c) and an expanded view of the same phonons in the RB₃ (d). The vertical lines used in b) and d) correspond to the same phonons for equivalent line styles (i.e. solid, dashed, and dotted). The solid (dashed) line spectra is the linearly polarized reflection taken along the [100] ([001]) crystallographic axis.

## Section S4: *Ab Initio* IR-active phonon modes and linewidths

In the phonon dispersion of α-MoO₃ (Figure S1.1), the IR-active TO modes that form the three MIR Reststrahlen bands (RB$_{1-3}$) reside in the O-dominated portion of the optical manifold above $\approx 500\,\text{cm}^{-1}$, consistent with the eigenvector projections in Figure S1.2 and the real-space TO eigen-displacements shown in Figure S1.3. For each isotopic sample, we extract the Γ-point TO and LO phonon frequencies and the corresponding TO linewidths from the DFPT calculated within the harmonic and three-phonon scattering formalisms mentioned previously (**Table S4.1-4**). These tables list the *ab initio* TO LO frequencies and linewidths that underlie the dielectric-function discussion and transfer-matrix modeling in the main text. The damping for each of the principal RB's TO phonon discussed in this work were plotted as a function of isotopic enrichment in **Figure S4.1**. Similar to the Raman linewidths (Figure 1b-d), we predict a systematic reduction in the IR-active TO phonon scattering linewidths, particularly for RB$_1$ and RB$_2$, due to $^{18}$O enrichment from first principles calculations, whereas $^{98}$Mo enrichment produces only minor changes. Precisely, we calculate the percent linewidth reduction between the averages of $^{16}$O and $^{18}$O isotopes resulting in 11.5%, 16.8%, and 13.2% for RB$_{1-3}$, respectively. These trends translate into modest lifetime improvements for the mid- and high-frequency RBs that drive the observed increase in HPhP Q-factors for $^{18}$O-enriched samples in the main text.

**Table S4.1:** *Ab Initio* dielectric function parameters of naturally abundant α-MoO₃.

| Main Axis | Mode Index | $\omega_{ij}^{TO}[cm^{-1}]$ | $\omega_{ij}^{LO}[cm^{-1}]$ | $\gamma_{ij}^{TO}[cm^{-1}]$ | $\gamma_{ij}^{LO}[cm^{-1}]$ |
|---|---|---|---|---|---|
| x | 1 | 471.359 | 488.288 | 22.414 | 26.373 |
| x | 2 | 764.376 | 938.595 | 7.715 | 8.484 |
| x | 3 | 1004.75 | 1005.319 | 0.274 | 0.258 |
| y | 1 | 532.123 | 825.976 | 22.602 | 12.467 |
| z | - | 759.477 | 766.715 | 10.547 | 8.431 |
| z | 1 | 963.34 | 1015.248 | 0.986 | 0.518 |

**Table S4.2:** *Ab Initio* dielectric function parameters of $^{98}$MoO₃

| Main Axis | Mode Index | $\omega_{ij}^{TO}[cm^{-1}]$ | $\omega_{ij}^{LO}[cm^{-1}]$ | $\gamma_{ij}^{TO}[cm^{-1}]$ | $\gamma_{ij}^{LO}[cm^{-1}]$ |
|---|---|---|---|---|---|
| x | 1 | 470.621 | 487.514 | 22.195 | 26.666 |
| x | 2 | 763.632 | 937.219 | 7.856 | 8.431 |
| x | 3 | 1003.222 | 1003.756 | 0.275 | 0.256 |
| y | 1 | 531.889 | 824.667 | 22.721 | 13.535 |
| z | - | 758.727 | 765.988 | 10.461 | 8.528 |
| z | 1 | 962.206 | 1013.71 | 0.996 | 0.509 |

**Table S4.3:** *Ab Initio* dielectric function parameters of $Mo^{18}O_3$

| Main Axis | Mode Index | $\omega_{ij}^{TO}[cm^{-1}]$ | $\omega_{ij}^{LO}[cm^{-1}]$ | $\gamma_{ij}^{TO}[cm^{-1}]$ | $\gamma_{ij}^{LO}[cm^{-1}]$ |
|---|---|---|---|---|---|
| x | 1 | 448.681 | 464.836 | 19.291 | 26.837 |
| x | 2 | 725.098 | 893.035 | 6.434 | 7.288 |
| x | 3 | 956.355 | 957.117 | 0.25 | 0.257 |
| y | 1 | 503.14 | 786.476 | 20.275 | 11.043 |
| z | - | 720.524 | 727.208 | 8.666 | 7.648 |
| z | 1 | 915.012 | 966.313 | 0.855 | 0.531 |

**Table S4.4:** *Ab Initio* dielectric function parameters of $^{98}Mo^{18}O_3$

| Main Axis | Mode Index | $\omega_{ij}^{TO}[cm^{-1}]$ | $\omega_{ij}^{LO}[cm^{-1}]$ | $\gamma_{ij}^{TO}[cm^{-1}]$ | $\gamma_{ij}^{LO}[cm^{-1}]$ |
|---|---|---|---|---|---|
| x | 1 | 447.926 | 464.048 | 19.024 | 26.869 |
| x | 2 | 724.31 | 891.587 | 6.52 | 7.456 |
| x | 3 | 954.728 | 955.448 | 0.249 | 0.252 |
| y | 1 | 502.887 | 785.091 | 19.818 | 11.195 |
| z | - | 719.727 | 726.437 | 8.725 | 7.733 |
| z | 1 | 913.803 | 964.676 | 0.866 | 0.522 |

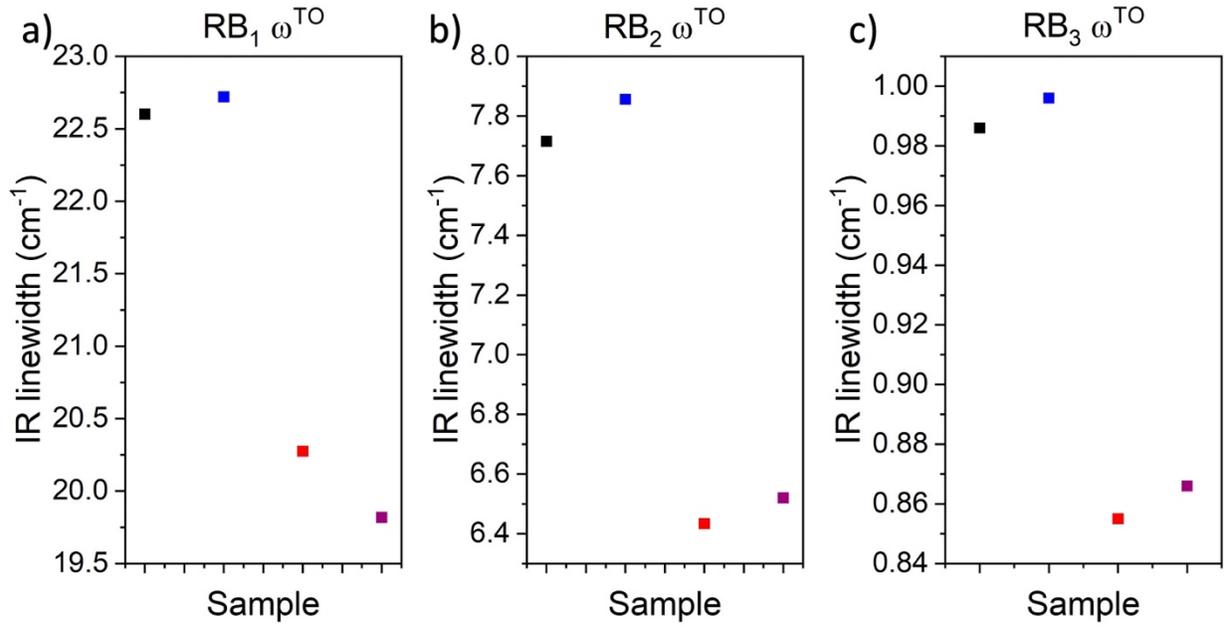

**Figure S4.1:** *Ab Initio* IR-active TO linewidths for each Reststrahlen band across isotopes (a-c).

## Section S5: TDTR measurements

Section S5.1: Cross-plane thermal conductivity measurements

To measure the cross-plane thermal conductivity of the MoO$_3$ flakes we use time-domain thermoreflectance (TDTR). In our TDTR setup, a Ti:sapphire laser with a central wavelength of ~800 nm and a repetition rate of 80 MHz emanates a train of sub-picosecond laser pulses which are split into a pump and a probe path. We modulate the 800 nm pump beam using an electro-optical modulator (EOM) at a frequency of 8.4 MHz. Using a lock-in amplifier and a balanced photodetector, the probe laser detects the reflectivity change due to surface temperature oscillations induced from the modulated pump train and, through use of a mechanical delay stage, measures the temperature decay up to 5.5 ns. These measurements implement a 10x objective which imposes 1/e$^2$ beam diameters of 11.7 µm and 19.3 µm for the probe and pump respectively.

The flakes are coated with an aluminum (Al) transducer in order to convert the optical energy to thermal energy within our sample stack. We use a two layer model to the analytical solution for the radially symmetric heat diffusion model to determine the cross-plane thermal conductivity of MoO$_3$ fitting for the thermal boundary conductance between Al and α-MoO$_3$ as well as the cross-plane thermal conductivity of α-MoO$_3$.[3,4] Our fitting results, shown in **Table S5.1**, indicate a constant thermal conductivity within uncertainty across the different isotopic enrichments. **Figure S5.1** exhibits an example best-fit of the thermal model compared to the ratio of the in-phase and out-of-phase data (–X/Y) for the $^{98}$MoO$_3$ enriched film. The parameters for our thermal model are depicted in **Table S5.2**.

**Table S5.1:** Cross-plane thermal conductivity and thermal boundary conductance fitting results for the α-MoO$_3$ films.

|  | $\kappa$ (W m$^{-1}$ K$^{-1}$) | Al-MoO$_3$ TBC (MW m$^{-2}$ K$^{-1}$) | TBC lower bound (MW m$^{-2}$ K$^{-1}$) | TBC upper bound (MW m$^{-2}$ K$^{-1}$) |
|---|---|---|---|---|
| Nat-MoO$_3$ | 2.03 ± .36 | 58.3 | 40.0 | 76.0 |
| Mo$^{18}$O$_3$ | 2.08 ± .37 | 52.6 | 37.5 | 75.4 |
| $^{98}$MoO$_3$ | 2.16 ± .37 | 55.3 | 40.6 | 81.5 |
| $^{98}$Mo$^{18}$O$_3$ | 2.06 ± .36 | 53.9 | 37.7 | 76.0 |

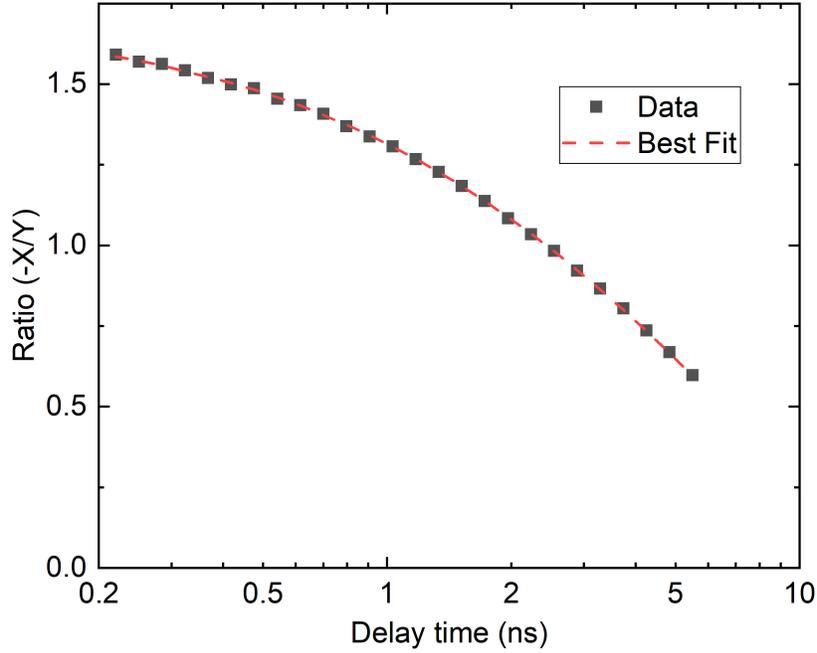

**Figure S5.1:** Example TDTR fit for the $^{98}$MoO$_3$ enriched film.

**Table S5.2:** Parameters used in the thermal model to determine the cross-plane thermal conductivity of α-MoO$_3$. The aluminum thermal conductivity and thickness were measured via 4-point probe and picosecond acoustics respectively.[5] We assume the α-MoO$_3$ as a semi-infinite substrate due to the large thickness of the flakes ~10 μm.

|  | Thermal conductivity W m$^{-1}$ K$^{-1}$ | Heat capacity, MJ m$^{-3}$ K$^{-1}$ | Thickness (nm) |
|---|---|---|---|
| Al | 144 (Measured) | 2.43 (Ref [S5]) | 91 (Measured) |
| α-MoO$_3$ | Fit | 2.44 (Ref [S6]) | - |

The uncertainty in our TDTR measurements is calculated using **Eq. S5.1** which accounts for spot-to-spot deviations as well as uncertainty in the fitting assumptions.[8]

$$\Delta = \sqrt{(\sigma^2) + (\sum_i \Delta_i^2) + (\sigma_C^2)} \qquad [\text{Eq.S5.1}]$$

where $\Delta$ is the total uncertainty, $\sigma$ is the standard deviation among multiple measurements across different spots, $\Delta_i$ is the uncertainty due to an individual parameter, and $\sigma_c$ is the contour uncertainty due to fitting assumptions[9,10]. For our uncertainty we assume a 5% uncertainty in aluminum heat capacity, aluminum thermal conductivity, and the thickness of the aluminum transducer. This range of uncertainty is typical in most TDTR measurements.[8] We calculate the uncertainty due to our fitting procedure via the method outlined by Feser *et al*.[11] comparing the residual between our fit and the experimental data for the fitted thermal parameters in our model. This allows for a better estimation of the uncertainty in the Al-MoO$_3$ TBC, which has much lower sensitivity than the α-MoO$_3$ thermal conductivity due to it being a lower resistance. For all of our measurements, our fitting procedure has low residuals around 0.005. Thus, for these measurements, we use a residual threshold of 0.01 added to the best fit residual to bound our fit indicating that any fit with a residual under ~ 0.015 is acceptable and contributes to our uncertainty.

Section S5.2: In-plane thermal conductivity measurements

For our in-plane thermal conductivity measurement we use a modified procedure to our cross-plane TDTR measurements to maximize the sensitivity to the in-plane thermal conductivity of α-MoO$_3$. First, we utilize a transducer of 15 nm Al/80 nm Ti, where the low thermal conductivity of titanium increases sensitivity to in-plane heat spreading in MoO$_3$ while the 15 nm Al surface layer allows us to maintain good thermoreflectance at our probe wavelength[12]. This transducer has been utilized in previous works to maximize sensitivity to in-plane thermal conductivity.[12] To further our increase our sensitivity to the in-plane heat spreading we use a 20 x objective (4.4 x 4.4 μm 1/e$^2$ diameter spot size) and low modulation frequency measurements at 1.2 MHz. This procedure of small spot size and low modulation frequency measurements to maximize sensitivity to in-plane thermal conductivity has also been described in previous works.[12,13] For these measurements we utilize a three-layer model incorporating both layers of the transducer and the α-MoO$_3$. We measure the properties of the transducer by performing 8.4 MHz modulation frequency measurements on two calibration samples with known thermal properties: Al$_2$O$_3$ and a-SiO$_2$, fitting for the thermal conductivity of the aluminum and titanium. For our model we assume an Al/Ti thermal boundary conductance of 3 GW m$^{-2}$ K$^{-1}$ which has been used in previous works.[12,14–16] Next, we perform 8.4 MHz modulation frequency measurements on our α-MoO$_3$ samples fitting for the cross plane thermal conductivity of the α-MoO$_3$ and the Ti/MoO$_3$ thermal boundary conductance. Using these thermal properties, we fit solely for the radial thermal conductivity of the α-MoO$_3$ layer. The results with the assumed values for our sample stack depicted are in **Table S5.3**; the results for all of these measurements are given in **Table S5.4**. The uncertainty for each of the fitted values is calculated using **Eq. S5.1** following the procedure described for the cross-plane thermal conductivity uncertainty analysis.

**Table S5.3:** Parameters used in the thermal model to determine the in-plane thermal conductivity of α-MoO₃. We assume the α-MoO₃ as a semi-infinite substrate due to the large thickness of the flakes ~10 µm.

|  | $\kappa_{\parallel}$ (W m$^{-1}$ K$^{-1}$) | $\kappa_{\perp}$ (W m$^{-1}$ K$^{-1}$) | Heat capacity, MJ m$^{-3}$ K$^{-1}$ | Thickness (nm) | Thermal boundary conductance (W m$^{-2}$ K$^{-1}$) |
|---|---|---|---|---|---|
| Al | Isotropic | 118.9 (Measured) | 2.43 (Ref. [6]) | 15 | - |
| Al/Ti | - | - | - | - | 3e9 (Refs. [12,14–16]) |
| Ti | Isotropic | 14.63 (Measured) | 2.36 (Ref. [17]) | 88 | - |
| Ti/MoO₃ | - | - | - | - | Fit |
| α-MoO₃ | Fit | Fit | 2.44 (Ref. [7]) | - | - |

**Table S5.4:** In-plane thermal conductivity results for the α-MoO₃ films including other thermal parameters that were used in the fitting procedure.

|  | $\kappa_{\parallel}$ (W m$^{-1}$ K$^{-1}$) | $\kappa_{\perp}$ (W m$^{-1}$ K$^{-1}$) | Ti-MoO₃ TBC (MW m$^{-2}$ K$^{-1}$) | TBC lower bound (MW m$^{-2}$ K$^{-1}$) | TBC upper bound (MW m$^{-2}$ K$^{-1}$) |
|---|---|---|---|---|---|
| Nat-MoO₃ | 9.8 ± 2.7 | 2.22 ± .30 | 92.4 | 59.9 | 204.5 |
| Mo¹⁸O₃ | 9.9 ± .2.6 | 2 ± .29 | 81.1 | 47.6 | 161.6 |
| ⁹⁸MoO₃ | 10.4 ± 2.94 | 2.29 ± .34 | 97.9 | 55.3 | 214.8 |

## Section S6: *Ab Initio* thermal conductivity and isotope mass-variance effects

The lattice thermal conductivity tensor was evaluated by solving the linearized phonon Boltzmann transport equation (BTE) within the relaxation-time approximation (RTA),

$$\kappa_{\alpha\beta}(T) = \left(\frac{1}{NV}\right) \sum_{\mathbf{q},j} C_{\mathbf{q}j}\, v_{\mathbf{q}j,\alpha} \otimes v_{\mathbf{q}j,\beta}\, \tau_{\mathbf{q}j}(T),$$ [Eq.S6.1]

where $V$ is the crystal volume and $N$ is the number of q-points, and the sum runs over wavevectors $\mathbf{q}$ and phonon branches s. The mode heat capacity $C_{\mathbf{q}j}$ and group velocity components $v_{\mathbf{q}j,\alpha}$ are computed from the harmonic IFCs, while $\tau_{\mathbf{q}j}(T)$ is the phonon lifetime determined by the total scattering rate. All results in **Figures S6.1-3** were obtained with phono3py using the same second- and third-order interatomic force constants as described in the main Methods, and the linearized BTE was solved at $T = 300\,\text{K}$.

Isotope disorder is treated using the Tamura mass-variance model. For each atomic species $\zeta$, we define the mass-variance parameter

$$g_\zeta = \sum_i f_{i\zeta} \left(\frac{m_{i\zeta}}{\overline{m_\zeta}-1}\right)^2,$$ [Eq.S6.2]

where $f_{i\zeta}$ and $m_{i\zeta}$ are the fractional abundance and mass of isotope $i$ on sublattice $\zeta$, and $\overline{m_\zeta}$ is the isotope-averaged mass. The corresponding isotope-disorder scattering rate $\Gamma_{\mathbf{q}j}^{\text{iso}}$ is added to the intrinsic anharmonic rate $\Gamma_{\mathbf{q}j}^{\text{anh}}$ obtained from three-phonon processes, $\Gamma_{\mathbf{q}j}^{\text{tot}} = \Gamma_{\mathbf{q}j}^{\text{anh}} + \Gamma_{\mathbf{q}j}^{\text{iso}}$, with $\tau_{\mathbf{q}j} = 1/(2\Gamma_{\mathbf{q}j}^{\text{tot}})$. In all cases, the dynamical matrix is constructed using isotope-averaged atomic masses, while the harmonic and anharmonic IFCs are kept fixed. This procedure captures both the mass-induced softening of phonon frequencies (and associated changes in group velocities) and the additional elastic scattering from mass disorder for each isotope configuration.

**Figure S6.1** summarizes the resulting thermal conductivity tensor components $\kappa_{xx}$, $\kappa_{yy}$, and $\kappa_{zz}$ at $300\,\text{K}$ for unenriched, $^{98}$Mo-enriched, $^{18}$O-enriched, and doubly enriched samples, with and without isotope scattering. The calculated in-plane thermal conductivity for natural-MoO$_3$ $\kappa_\| \approx 10\,\text{Wm}^{-1}\text{K}^{-1}$ shows excellent quantitative agreement with our experimental TDTR measurements reported in Section S5 ($9.8 \pm 2.7\,\text{Wm}^{-1}\text{K}^{-1}$). This agreement validates the accuracy of the underlying harmonic and anharmonic force constants used here to decompose the thermal transport properties.

The strong anisotropy $\kappa_{xx} > \kappa_{yy} \gg \kappa_{zz}$ reflects the layered orthorhombic structure and is consistent with the TDTR measurements in the main text. Eliminating the Mo isotope disorder ($^{98}$MoO$_3$) produces only a modest increase in κ, whereas $^{18}$O enrichment lowers κ in all directions despite the removal of O-isotope scattering, demonstrating that the dominant effect of $^{18}$O is the

≈12.5% oxygen mass increase and corresponding reduction in phonon group velocities rather than additional elastic disorder.

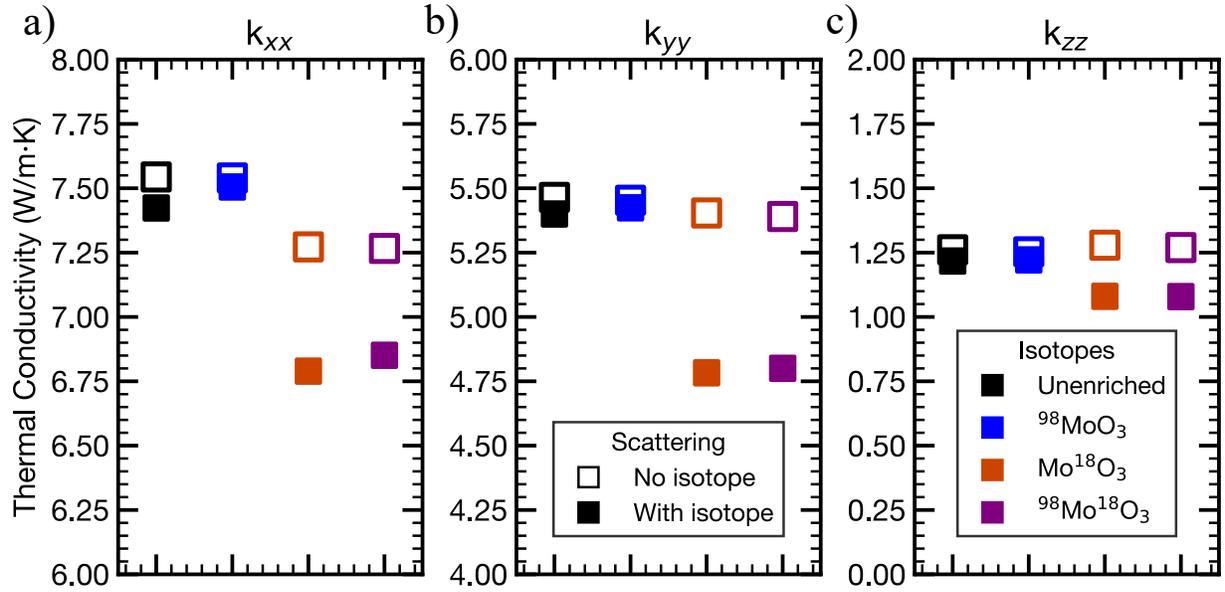

**Figure S6.1:** *Ab initio* thermal conductivity values in isotopically enriched along each principal crystallographic direction α-MoO$_3$. Calculated lattice thermal conductivity tensor components $\kappa_{xx}$, $\kappa_{yy}$, and $\kappa_{zz}$ at 300 K for unenriched (natural isotope abundance, black), $^{98}$Mo-enriched ($^{98}$MoO$_3$, blue), $^{18}$O-enriched (Mo$^{18}$O$_3$, red), and doubly enriched ($^{98}$Mo$^{18}$O$_3$, purple) samples. Open squares denote calculations including only three-phonon anharmonic scattering, while filled squares include both anharmonic and isotope scattering contributions.

The microscopic origin of this behavior is clarified in **Figure S6.2**, which decomposes κ into frequency-, momentum-, and species-resolved contributions, aided by the atom-projected phonon dispersion in Figure S1.2. At 300 K, acoustic modes (ω < 180 cm$^{-1}$) carry the majority of the heat current in all directions—about 68% of $\kappa_{xx}$, 60% of $\kappa_{yy}$, 70% of $\kappa_{zz}$, and 65% of $\kappa_\| =$ ($\kappa_{xx}+\kappa_{yy}$)/2—while optical modes contribute the remaining 30–40%. The in-plane conductivity is dominated by long-wavelength acoustic modes with mixed Mo–O character at small | **q** |/**q**$_{BZ}$, whereas $\kappa_{zz}$, though still primarily acoustic, receives a more broadly distributed contribution across momentum space, reflecting the weaker interlayer van der Waals bonding along z. The species-projection analysis further shows that modes with substantial participation from both Mo and O, rather than purely Mo- or purely O-like vibrations, are the main heat carriers in both $\kappa_\|$ and $\kappa_{zz}$, which explains their sensitivity to changes in the oxygen mass under $^{18}$O enrichment.

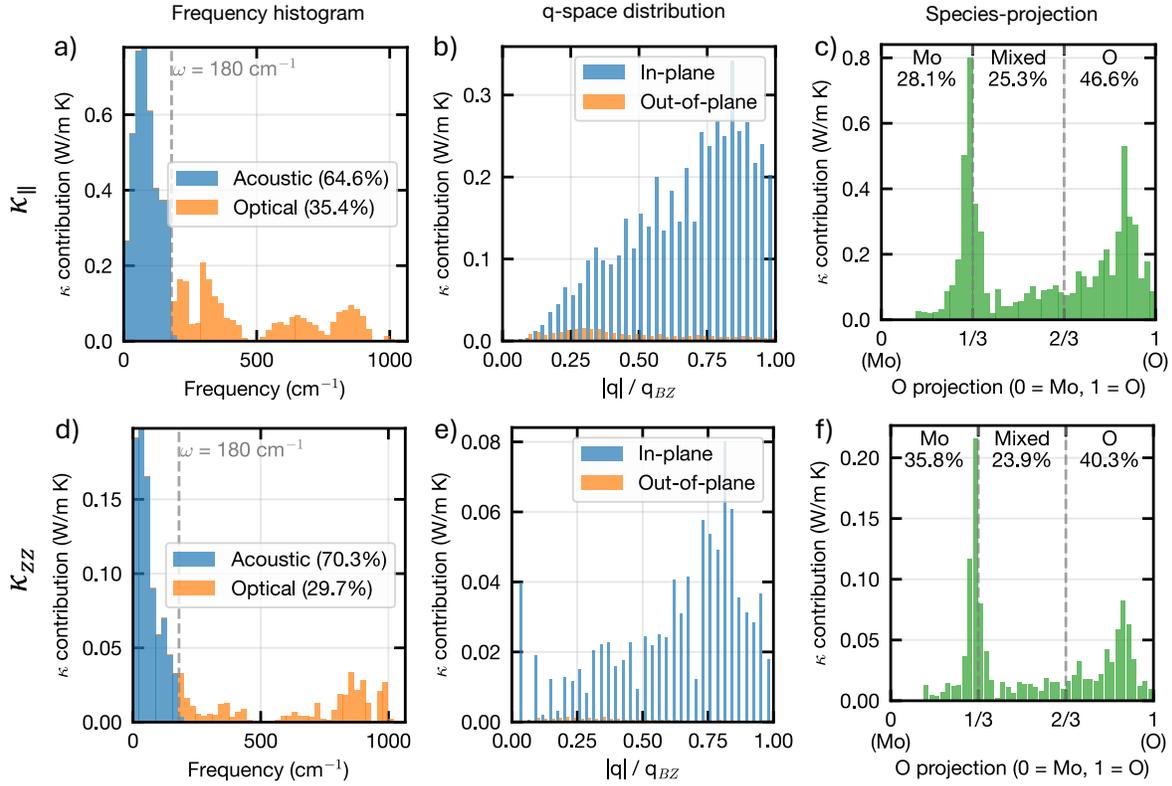

**Figure S6.2:** Mode-resolved decomposition of the calculated lattice thermal conductivity in α-MoO$_3$ at 300 K. (a, d) Frequency-resolved κ contributions showing acoustic (blue, ω < 180 cm$^{-1}$) and optical (orange, ω > 180 cm$^{-1}$) phonon contributions for in-plane $\kappa_\parallel = (\kappa_{xx} + \kappa_{yy})/2$ (a) and cross-plane $\kappa_{zz}$ (d). The acoustic/optical cutoff is marked by the dashed vertical line, consistent with the acoustic and optical bands in Figure S1.1. (b, e) Momentum-resolved κ contributions as a function of normalized wavevector magnitude $|q|/q_{BZ}$, decomposed by q-vector orientation into in-plane ($q_\parallel$, blue) and out-of-plane ($q_z$, orange) components for $\kappa_\parallel$ (b) and $\kappa_{zz}$ (e). (c, f) Species-resolved κ contributions as a function of oxygen character $P_O$, where $P_O$ is the squared eigenvector projection onto oxygen atoms summed over all O sites. Vertical dashed lines at $P_O = 1/3$ and $2/3$ delineate Mo-dominated ($P_O < 1/3$), mixed ($1/3 \leq P_O \leq 2/3$), and O-dominated ($P_O > 2/3$) character regions for $\kappa_\parallel$ (c) and $\kappa_{zz}$ (f). The analysis reveals that in-plane thermal transport is dominated by long-wavelength (small $|q|$) acoustic modes with mixed Mo–O character, while cross-plane transport shows a more distributed contribution across momentum space, reflecting the weak interlayer van der Waals bonding along the z-direction.

Finally, **Figure S6.3** recasts the κ changes under isotope enrichment in terms of their basic BTE ingredients via a component-substitution analysis. By recomputing κ while substituting, one at a time, the mode heat capacities, squared group velocities, anharmonic scattering rates, or isotope-scattering rates from the enriched cases into the unenriched reference (with $\Gamma^{iso} = 0$), we isolate their individual contributions to Δκ. The results show that $^{18}$O enrichment affects κ primarily through reduced group velocities, with only minor changes arising from heat capacity and intrinsic anharmonic scattering, while the explicit isotope-disorder term mainly reflects the removal of Mo mass variance in the $^{98}$Mo-enriched case and the predicted percent enrichment of $^{18}$O. Further investigation of the thermal conductivity along the [100] and [001] could demonstrate the preferential direction of in-plane heat transport and isotopic enrichment could further enhance that degree of anisotropy. The deterministic design of a twisted bilayer isotopic α-MoO$_3$ heterostructure could provide a platform of steerable and directional in-plane dissipation of heat. Figures S6.1-3 represent different reorganizations of the same underlying ab initio BTE dataset.

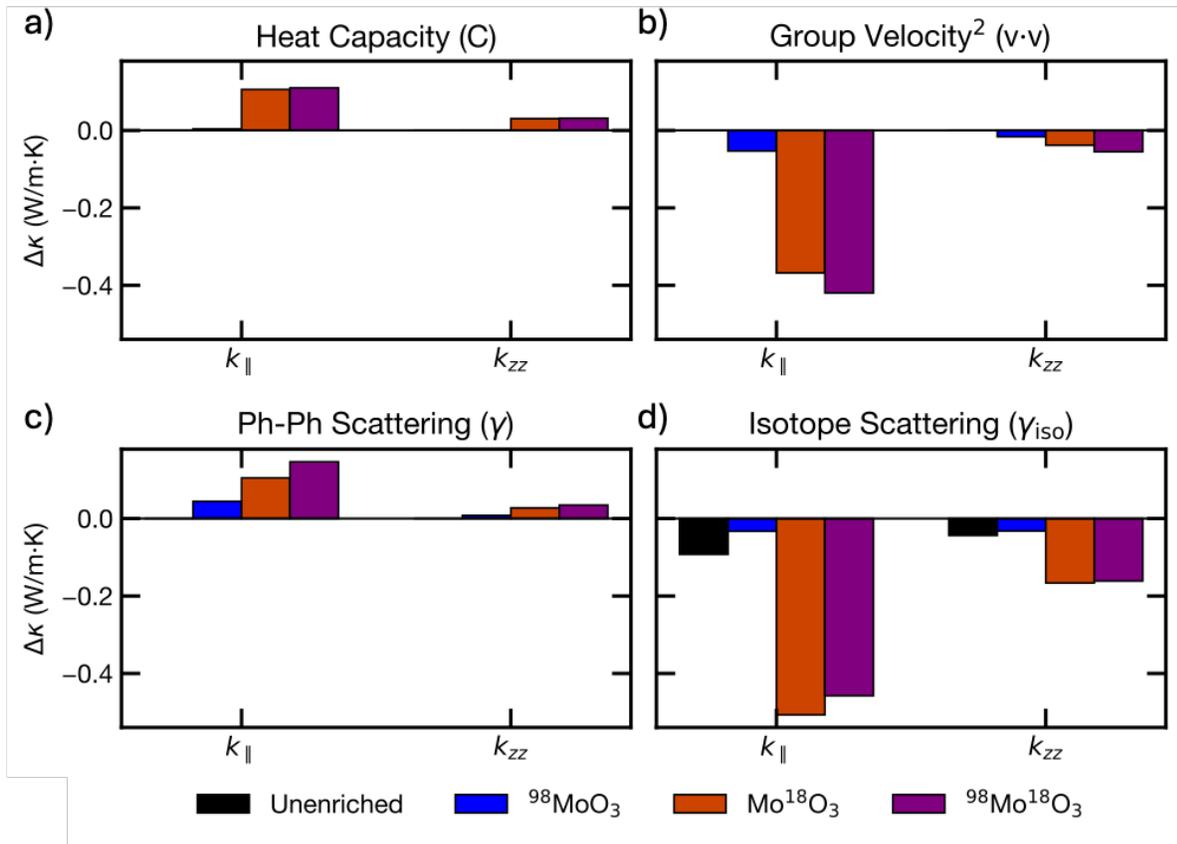

**Figure S6.3:** Decomposition of the calculated changes in thermal conductivity from isotopic enrichment into components of a) phonon heat capacity, b) group velocity, c) scattering rates, and d) isotopic disorder scattering rates. Component-substitution analysis — isolating individual contributions to the thermal conductivity — change κ, Δκ, upon isotope enrichment. Each panel shows the change in in-plane ($\kappa_\parallel = (\kappa_{xx} + \kappa_{yy})/2$) and cross-plane ($\kappa_{zz}$) thermal conductivity when only the specified component is substituted from the enriched case, with all other components held at their unenriched baseline values (with $\gamma_{iso} = 0$). Colors denote unenriched (black), $^{98}$Mo-enriched (blue), $^{18}$O-enriched (red), and doubly enriched (purple) configurations.

## Section S7: s-SNOM sample images and AFM measurements

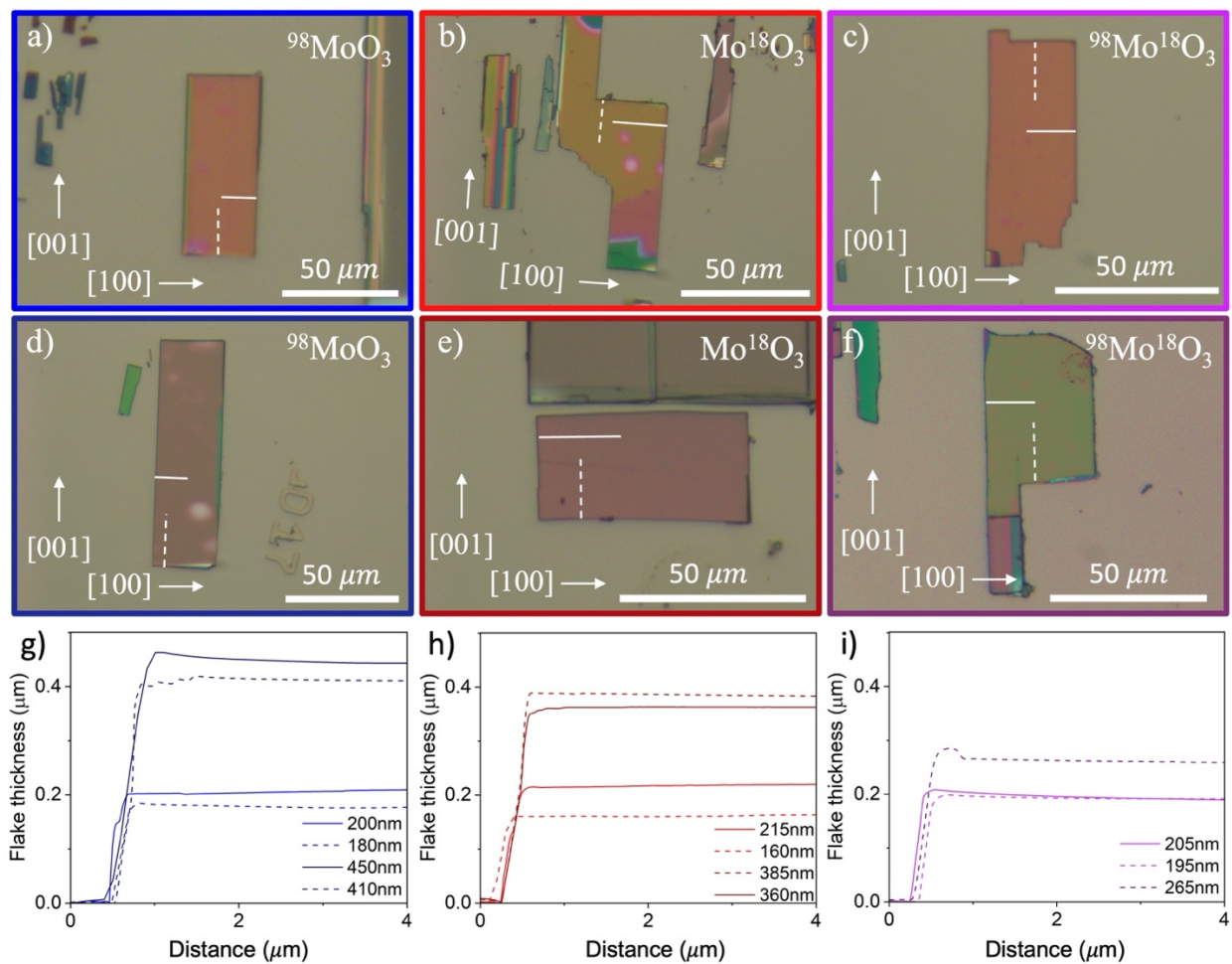

**Figure S7.1:** Visible light microscope images of the thinner (a-c) and thicker (d-f) set of isotopic α-MoO₃ flakes prepared for s-SNOM measurements with their respective thicknesses mapped from AFM and plotted in (g-i). The solid (dashed) line are line scans taken along the [100] ([001]) crystallographic direction. The darker color scheme (thicker flakes) corresponds to the samples in (d-f).

## Section S8: FFT analysis of HPhPs in real and momentum space

We extract the frequency dependent line profiles of HPhPs propagating in each isotopic slab of α-MoO$_3$ from the s-SNOM images. The beginning of the line scan is taken from the flake edge by mapping the AFM s-SNOM data and coinciding the same pixels in the near-field data to ensure that the profiles as a function of distance from the flake edge. In this section, we examine the line profile extracted from an HPhP excited at 930 cm$^{-1}$ propagating along the [001] in the RB$_3$ of a ~400 nm Mo$^{18}$O$_3$ flake. First and foremost, the raw HPhP profile is put through a high-frequency pass FFT filter to exclude any low frequencies present from the far-field excitation source (a bandpass FFT filter is optional to exclude very high frequencies from a low signal-to-noise ratio) and the DC offset is removed from the FFT. The resulting HPhP profile is seen in **Figure S8.1a**, where the line scan is symmetric about 0 amplitude, the far-field oscillations have vanished, and the near-field profile is still preserved. Aside from the additional oscillations in the first fringe, it is clear that both edge-launched ($\lambda_P$) and tip-launched ($\frac{\lambda_P}{2}$) HPhPs are present. Following the far-field correction, we scale the near-field amplitude by multiplying a geometric correction factor ($\sqrt{x}$) to account for the radial propagation of a point-launched HPhP. With the HPhP fully prepped for analysis, we apply an FFT with a Hanning window and minimal zero-padding to make the FFT peaks as close to symmetric as possible. The resulting Fourier spectra, F(x), in **Figure S8.1b** displays three distinct peaks to which we fit the following Lorentzian profile to each peak

$$F(k) = F_0 + \frac{2A}{\pi}\left(\frac{w}{4(k-k_c)^2 + w^2}\right) \qquad [\text{Eq.S8.1}]$$

where k is the wavevector, k$_c$ is the center wavevector of the fitted peak, $w$ is the full-width half maximum (FWHM) of the fitted peak, A is the area of the peak, and F$_0$ is the vertical offset. For each fitted peak, the complex wavevector of each modal order of HPhPs present can be extracted by $Re(k) = k_c$ and $Im(k) = FWHM$. Given by the relationship from edge or tip-launched HPhPs, the $Re(k)$ of a tip-launched HPhP is twice the frequency of the edge-launched counterpart. Such a frequency relationship is clear in **Figure S8.1b**, where the edge and tip-launched peaks are clearly identified with an additional higher frequency peak present. The free-space normalization of the first and third peaks are included in the data set plotted in **Figure 3e** at 930 cm$^{-1}$, confirming their identification as the *l*=0 and 1 edge-launched HPhPs. At this point, we calculate the Q-factors from the complex from the complex HPhP wavevector extracted from the Fourier spectra.

To ensure that Q-factors extracted from the Fourier spectra are accurate, we investigate the real and imaginary components independently and compare them with the real-space extracted values. The real-space counterparts are calculated from the Fourier spectra (inset values in

**Figure S8.1b**) to the polariton wavelength and propagation length by $\lambda_P = \frac{1}{Re(k)}$ and $L_P = \frac{1}{Im(k)}$, respectively. To compare these values to the direct extraction of the HPhP wavelength and propagation length, we take the scaled HPhP line profile prior to applying the FFT and instead, perform a FFT bandpass filter. Here, we selectively filter the line profile for either the *l*=0 or 1 order and plot them in **Figure S8.1c** and **Figure S8.1d**, respectively. For the *l*=0 mode, we filter both the edge and tip-launched frequencies together since their independent filtering led to an inconsistent frequency with respect to the raw line profile. Both FFT filtered HPhP profiles were fitted the damped sine function

$$S(\omega)\,\sigma_3 = S_0 + A\frac{e^{-\frac{2x}{L_P}}}{\sqrt{x}}\sin\left(\frac{4\pi(x-x_a)}{\lambda_P}\right) + B\frac{e^{-\frac{x}{L_P}}}{x}\sin\left(\frac{2\pi(x-x_b)}{\lambda_P}\right) \qquad [\text{Eq.S8.2}]$$

where A (B) is the amplitude of the tip(edge)-launched HPhP, $x_{a(b)}$ is the phase shift applied to the tip(edge)-launched HPhP, and $S_0$ is the vertical offset. For the *l*=0 mode, both A and B are non-zero, but we fix A=0 for the *l*=1 mode since we previously validated its wavevector as an edge-launched mode with the HPhP dispersion. We observe excellent agreement between both the direct extraction of the *l*=0 mode wavelength and propagation length and the values calculated from the Fourier spectra. Thus, we employ the Fourier analysis of the HPhP wavevectors due to the presence of multiple higher-order modes in our s-SNOM measurements. We acknowledge the discrepancy in the fit in **Figure S8.1c** which shows a degree of over damping for the edge launched but an under damping for the tip-launched HPhP. This is possibly due to the correction factor we apply to the data before fitting. We provide the real-space evaluation of the *l*=1 mode here to demonstrate the challenges with reporting propagation lengths or Q-factors of higher-order modes. Their damping is likely underestimated due to the dominance of the fundamental mode upon the fringes that we can resolve from the higher-order modes. Therefore, we can only accurately report the $Re(k)$ of the higher-order modes.

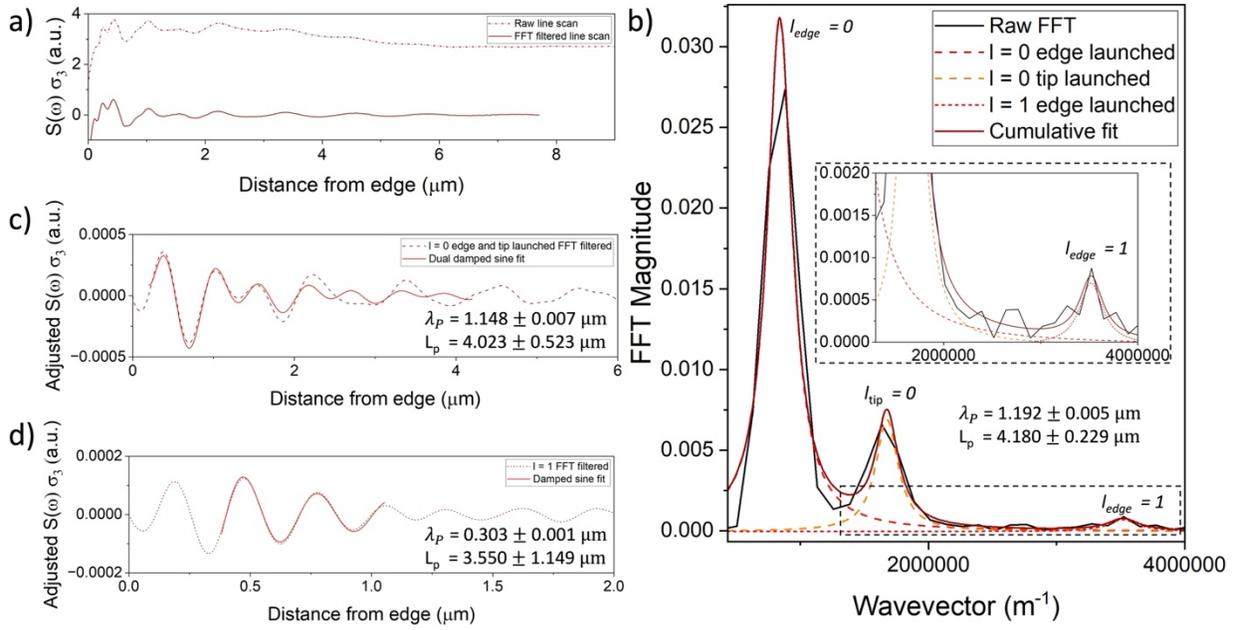

**Figure S8.1:** FFT analysis to extract complex HPhP wavevector from real space and momentum space.

## Section S9: Additional dispersion mapping for various thicknesses

Here we provide the remaining experimentally mapped dispersion points for the isotopically enriched α-MoO$_3$ that was not provided in the main text. As seen with Figure 3, we observe excellent agreement between the TMM calculations and the HPhP wavevector extracted from s-SNOM measurements.

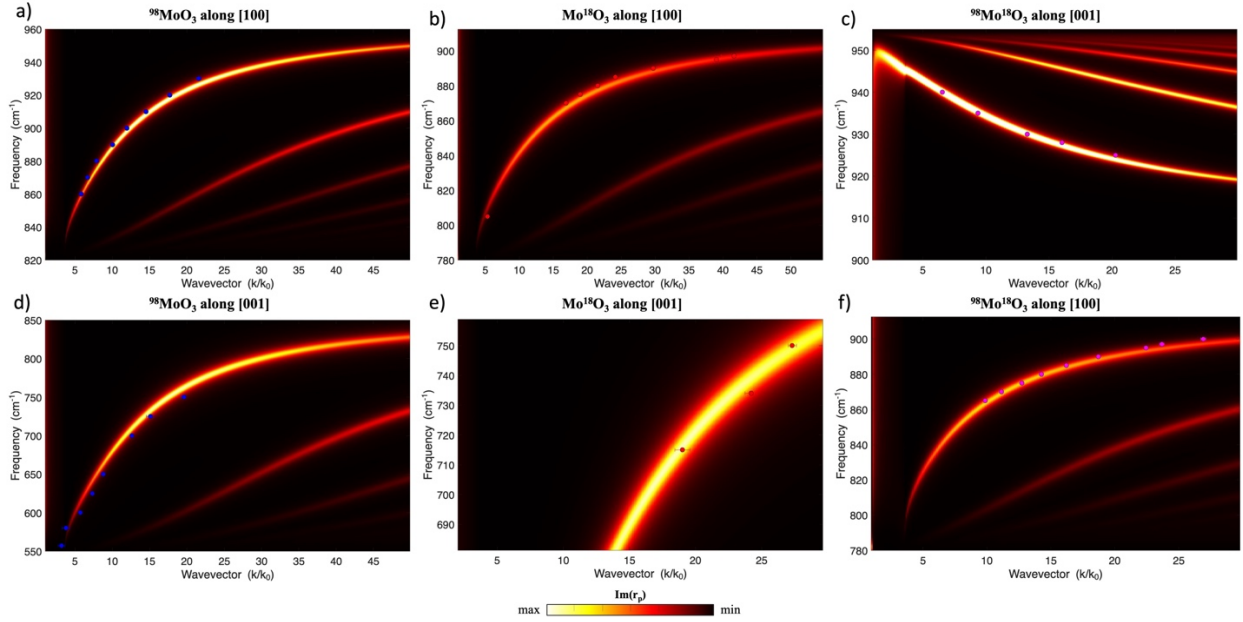

**Figure S9.1:** Experimental dispersion points mapped on TMM calculations of p-polarized Im($r_p$) in a) ~200nm $^{98}$MoO$_3$ along the [100], b) ~200nm Mo$^{18}$O$_3$ along the [100], c) ~300nm $^{98}$Mo$^{18}$O$_3$ along the [001], d) ~200nm $^{98}$MoO$_3$ along the [001], e) ~200nm Mo$^{18}$O$_3$ along the [001], and f) ~300nm $^{98}$Mo$^{18}$O$_3$ along the [100].

### Section S10. Comparison in dielectric function with literature

Prior characterizations of $^{92}$MoO$_3$ and $^{100}$MoO$_3$ has been performed by *Schultz et al.*[18] and *Zhao et al.*[19], with their findings provided in **Table S10.1**. For reference to naturally abundant α-MoO$_3$, we also include the parameters reported by *Álvarez-Pérez et al.*[2] The high-frequency dielectric constants for the isotopes in this work were manually tuned to fit the higher-order HPhP branches with our experimental data on the $l=2$ and $l=3$ branches. The resulting summary provides the full spectral tunability available in α-MoO$_3$ through selective isotopic enrichment.

**Table S10.1: Summary of dielectric function tunability in isotopically enriched α-MoO$_3$**

| Isotope | Source | $\omega_x^{TO}$ (cm$^{-1}$) | $\omega_x^{LO}$ (cm$^{-1}$) | $\varepsilon_x^\infty$ | $\omega_y^{TO}$ (cm$^{-1}$) | $\omega_y^{LO}$ (cm$^{-1}$) | $\varepsilon_y^\infty$ | $\omega_z^{TO}$ (cm$^{-1}$) | $\omega_z^{LO}$ (cm$^{-1}$) | $\varepsilon_z^\infty$ |
|---|---|---|---|---|---|---|---|---|---|---|
| $^{92}$MoO$_3$ | Ref [S3] | 820 | 978 | 4.7 | 545 | 851 | 5.0 | 964 | 1008 | 2.6 |
| $^{92}$MoO$_3$ | Ref [S2] | 822 | 976 | 3.7 | – | – | – | 963 | 1008 | 2.3 |
| $^{98}$MoO$_3$ | This work | 824 | 966 | 4.3 | 544 | 849 | 7.9 | 960 | 1009 | 2.0 |
| $^{100}$MoO$_3$ | Ref [S3] | 816 | 975 | 4.2 | 545 | 851 | 5.0 | 956 | 1002.5 | 3.0 |
| $^{100}$MoO$_3$ | Ref [S2] | 816 | 974 | 4.2 | – | – | – | 956 | 1002.5 | 3.0 |
| Nat-MoO$_3$ | Ref [S1] | 821.4 | 963.0 | 5.78 | 544.6 | 850.1 | 6.59 | 957.6 | 1006.9 | 4.47 |
| Mo$^{18}$O$_3$ | This work | 779 | 923 | 5.8 | 526 | 806 | 5.1 | 912 | 954 | 4.5 |
| $^{98}$Mo$^{18}$O$_3$ | This work | 779 | 923 | 5.8 | 526 | 806 | 6.1 | 910 | 954 | 3.6 |

## Section S11: Additional HPhP Q-factor plots

The momenta of HPhPs supported in thicker flakes of isotopically enriched α-MoO$_3$ were also fully analyzed. As discussed in the main text, the experimental challenges present to accurately compare spectrally shifted HPhPs in the RB$_1$ due to isotopic enrichment are present in this set of data (**Figure S11.1a**). With further inspection between **Figure 4a** and **Figure S11.1a**, the Q-factors between $^{98}$MoO$_3$ in the main text and Mo$^{18}$O$_3$ are nearly identical in the low wavevector regime. Despite these two samples being significantly distinct in thickness, they were taken during the same day and suffer the same degree of absorption loss from the RB$_1$ TO phonon. Therefore, we draw the connection that the RB$_1$ TO phonon experiences minimal lifetime improvements from $^{18}$O enrichment and is in agreement with the prediction of the degree is restriction upon the O(1) atom discussed in Section S1. Within the RB$_2$, we observe offset maxima for each isotopic sample which is likely due to the larger discrepancy of sample thickness where the normalized wavevector, k*d, only holds for comparison in small discrepancies of sample thickness[18] (**Figure S11.1b**). However, we still see the improvements in Q-factor in the high-wavevector regime dominated by the expected lifetime improvements due to $^{18}$O enrichment. As seen in both the RB$_2$ and RB$_3$, the Q-factors decrease for thicker samples which can be understood through additional scattering losses present as the thickness of the flake is increased (**Figure S11.1c**).

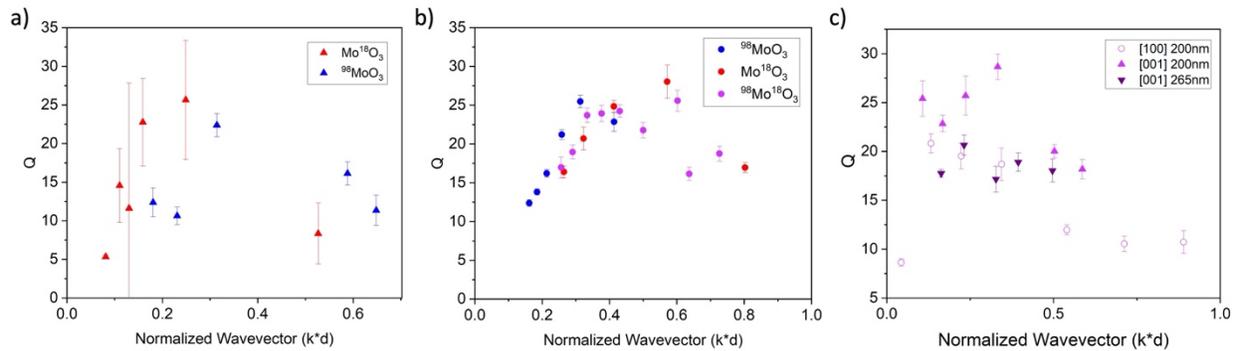

**Figure S11.1:** HPhP Q-factors of thicker isotopically enriched α-MoO$_3$ flakes in the RB$_1$, RB$_2$, and RB$_3$ for a-c), respectively. Note that the only data set which includes any Q-factors from the thinner flakes is in c).

Here we provide a brief reporting of HPhP Q-factors propagating along the [001] within the RB$_1$ of a 200 nm naturally abundant α-MoO$_3$ flake shown in **Figure S11.2**. Unlike the RB$_2$ and RB$_3$, which naturally abundant α-MoO$_3$ Q-factors has been previously reported, the RB$_1$ has failed to be investigated. Thus, we report these values as a comparison to our isotopic Q-factors. As a reminder, $^{98}$MoO$_3$ is selected in this study to serve a close representation of the mass of naturally abundant α-MoO$_3$ without the $^X$Mo isotopic disorder.

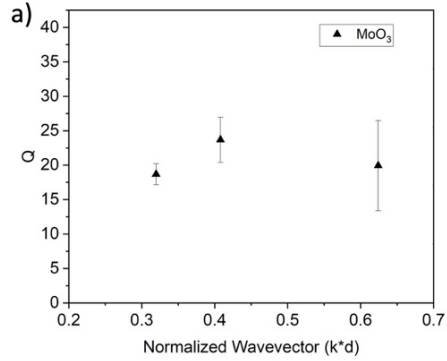

**Figure S11.2:** RB$_1$ HPhP Q-factors extracted from a 200 nm naturally abundant α-MoO$_3$ flake.

### Section S12: Free space confinement from $^{18}$O enrichment

Aside from the Q-factor enhancements originating from the imaginary HPhP wavevector, we also investigate the changes in the real-component HPhP wavevector. Due to the thickness dependence of the complex wavevector shown in Equation 3 of the main text and the invalidity for large thickness discrepancies discussed in Section S11, we restrict this investigation to the data set of samples with comparable thicknesses in which we observe a Q-factor enhancement. Within the RB$_2$ of our thinner set of isotopic flakes, we first plot the free space normalized wavevector as a function of the wavenumber differential from the TO phonon defined as

$$\Delta\omega = \omega - \omega^{TO} \qquad [Eq.S12.1]$$

where ω is the excitation wavenumber and $\omega^{TO}$ is the TO phonon wavenumber. There is a clear difference between $^{16}$O and $^{18}$O flakes in terms of how much their HPhPs are compressing the free-space wavevector as seen in **Figure S12.1a**. While this relationship accounts for the redshift in the TO phonon frequency from $^{18}$O enrichment, the slight discrepancy in thickness makes for an unjust comparison. Furthermore, for an equal redshift in phonon frequencies between $^{98}$Mo$^{18}$O$_3$ and Mo$^{18}$O$_3$, we expect an equal level of confinement but do not observe as such. Since the only distinction between these two isotopes are additional scattering losses which alter the imaginary wavevector component, we conclude these slight differences in confinement are a result of the slight thickness discrepancies skewing this inverted dispersion relationship. This further supports the reasoning discussed in the main text as to why the Q-factors could not be compared in any form of frequency. Therefore, we plot the free-space confinement factor with respect to the thickness normalized wavevector (**Figure S12.1b**). While this is a straightforward representation of free-space confinement, we eliminate any concerns from the dispersion dependence upon thickness. As a result, we still observe a higher degree of confinement in the $^{18}$O flakes over the $^{16}$O case (**Figure S12.1c**). Importantly, the increase in confinement is equal across both $^{18}$O flakes, confirming the expectations described earlier. The implications of this increase in confinement is best understood when considering wavelengths, which are inversely proportional to wavevector. Thus, the HPhPs in the $^{18}$O flakes achieve higher compression because they achieve the same polariton wavelength for a larger excitation wavelength due to the red shift from $^{18}$O enrichment.

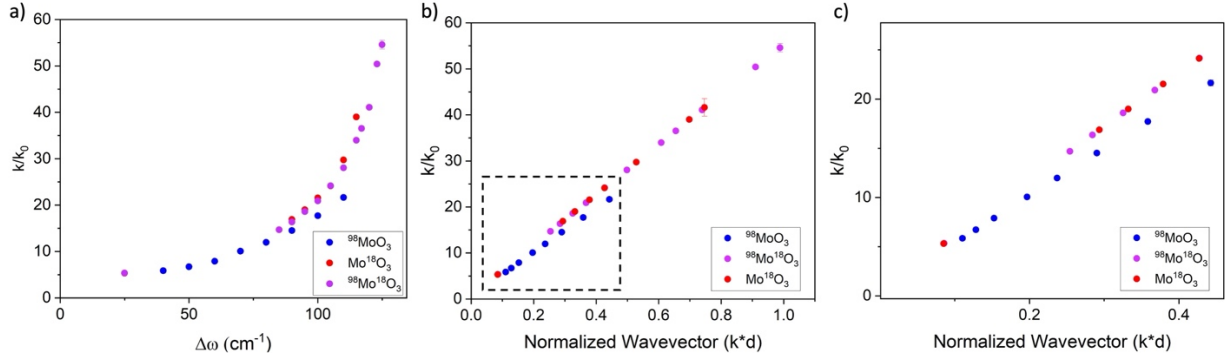

**Figure S12.1:** HPhP confinement of free-space light in the RB$_2$ as a function of the difference in wavenumber from the TO phonon (a) and thickness normalized wavevector (b and c); where c) is zoomed into the region comparing all three isotopes.

## S13. Individual RB$_3$ Q-factors in isotopically enriched α-MoO$_3$ flakes

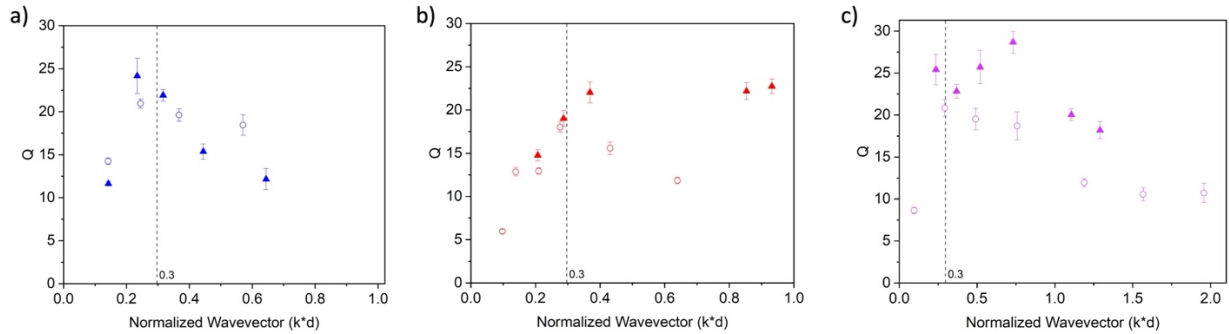

**Figure S13.1:** Isotopic HPhP Q-factors within the RB$_3$ shown in Fig.4c of the main text plotted in separately for clarity; where a) is $^{98}$MoO$_3$, b) is Mo$^{18}$O$_3$, and c) is $^{98}$Mo$^{18}$O$_3$.